\documentclass[sigplan,review,nonacm, 10pt]{acmart}
\renewcommand\footnotetextcopyrightpermission[1]{}
\PassOptionsToPackage{hyphens}{url}
\usepackage[shortcuts]{extdash}  
\usepackage{multicol} 
\usepackage{tabularx, url} %
\usepackage{graphicx} 
\usepackage{enumitem}
\usepackage{hyperref}
\usepackage{xcolor}
\usepackage[many]{tcolorbox}
\usepackage{makecell}
\usepackage{xspace}
\usepackage{amsmath}
\usepackage{xparse}
\usepackage{array}
\usepackage{listings}
\definecolor{Purple}{HTML}{A020F0}
\usepackage{fancyvrb}
\usepackage{verbatimbox,caption,float,lipsum}
\usepackage{multirow}
\usepackage[labelfont=bf,font=small]{subcaption}

\usepackage{amssymb}
\usepackage[table]{xcolor} 
\usepackage{threeparttable}

\usepackage[capitalize, nameinlink]{cleveref}
\crefformat{section}{#2\S{}#1#3}
\Crefname{section}{Section}{Sections}
\makeatletter
\Crefformat{section}{\Cref@section@name~#2#1#3}
\makeatother

\usepackage{tikz}
\newcommand{\cnumber}[1]{\tikz[baseline=(myanchor.base)] \node[circle,fill=.,inner sep=1pt] (myanchor) {\color{-.}\bfseries\footnotesize #1};}

\newcolumntype{P}[1]{>{\centering\arraybackslash}p{#1}}
\newcolumntype{M}[1]{>{\centering\arraybackslash}m{#1}}
\newcolumntype{L}[1]{>{\raggedright\arraybackslash}m{#1}}

\definecolor{darkblue}{RGB}{0, 51, 102}
\definecolor{mygreen}{RGB}{0, 150, 0}
\definecolor{myred}{RGB}{200, 0, 0}

\newcommand{\greencheck}{\textcolor{mygreen}{\checkmark}}
\newcommand{\redcross}{\textcolor{myred}{$\times$}}

\newcommand{\sys}{\textsc{TEEVisor}\xspace}

\begin{document}

\date{}

\title{\Large \bf Enabling Extended SGX Enclaves inside Confidential Virtual Machines}

\author{Pamenas Kariuki}
\affiliation{%
  \institution{TU Dresden}
  \country{Germany}
}
\author{Dong Yiliang}
\affiliation{%
  \institution{TU Dresden}
  \country{Germany}
}
\author{Quoc Do Le}\authornote{Quoc Do Le performed this work at TU Dresden.}
\affiliation{%
  \institution{StackIT Cloud}
  \country{Germany}
}
\author{Christof Fetzer}
\affiliation{%
  \institution{TU Dresden}
  \country{Germany}
}

\pagestyle{plain}
\begin{abstract}

Confidential Virtual Machines (CVMs), such as AMD SEV-SNP, and process enclaves, such as Intel SGX, provide workload protection against cloud threats. However, they strongly differ in TCB, deployment cost and speed, peer workload protection, resource sharing, fine-grained isolation and attestation, and programming capabilities. Nesting process enclaves within CVMs offers an optimal solution and an opportunity to address long-standing SGX limitations: lack of intra-enclave protection, expensive transitions and exception paths, slow enclave cloning, and a restricted programming model that limits Library Operating Systems (LibOS).

This paper introduces \sys{}, a CVM-resident architecture consisting of a security monitor and a kernel driver that leverages AMD SEV-SNP Virtual Machine Privilege Levels (VMPLs) to build hardware-protected, process-level enclaves inside a guest CVM with two modes: (1) an SGX binary compatibility mode to run unmodified SGX applications; and (2) a novel extended SGX~(eSGX) mode optimizing slow execution paths and offering richer enclave programming capabilities. On commercial AMD EPYC hardware, against native CVM execution, \sys{} incurs under 5\% overhead on compute- and memory-bound workloads, around 12\% on I/O-heavy applications, and accelerates enclave cloning to a rate of 93~$\mu$s/MB of enclave size, demonstrating that eSGX semantics inside a CVM are practical for production use.

\end{abstract}
\maketitle
\thispagestyle{empty}

\section{Introduction}

Confidential computing has become a cornerstone primitive for running sensitive cloud workloads on shared infrastructure. Two common Trusted Execution Environment (TEE) abstractions are: \emph{process-based enclaves} such as Intel SGX~\cite{costan_intel_2016}, isolating userspace components in behind a small Trusted Computing Base (TCB), and \emph{confidential virtual machines} (CVMs) such as AMD SEV-SNP~\cite{kaplan_amd_nodate-1} and Intel TDX~\cite{al_intel_2023}, which protect a VM from a malicious hypervisor. SGX is prevalent in server CPUs and at the core of Intel TDX attestation.

Critically, neither enclaves nor CVMs alone meet the demands of modern cloud services: enclaves retain a small, auditable TCB but impose rigid memory layouts, expensive transitions, and a poor programming model; CVMs offer binary compatibility and rich OS services but expose a large attack surface~\cite{duy_incognitos_2025,kuvaiskii_gramine-tdx_2024}.

This tension has concrete operational consequences. Workload isolation, resource sharing, deployment cost, and TCB size constrain confidential cloud orchestration: OS-level containers CVM offer limited protection against a compromised guest OS, while a single workload per CVM is costly, slow to provision, hard to scale, and weakens per‑workload attestation. MicroCVM isolation (e.g., unikernels) reduce TCB but has same scaling issues. Some domains have strict requirements (e.g., healthcare~\cite{gematik_whitepaper_2025}) for confidential isolation of workloads, and demand either enclaves or exclusive CVM per workload that is costlier or has higher orchestration complexity. Thus, we are motivated to explore solutions at the intersection of CVMs and process-based enclaves.

Architectural bottlenecks in SGX have hindered its broader adoption: enclave exceptions force costly transitions and host interaction, degrading memory management and latency in exception-heavy applications (JVM~\cite{zheng_empirical_2017}); host-mediated signaling and transitions for preemption; and emulating \texttt{fork}/\texttt{clone} using sealing and state copying has significant overheads despite SGXv2 improvements and runtimes like Haven~\cite{haven_2015}, SCONE~\cite{scone}, Graphene‑SGX~\cite{tsai_graphene-sgx_2017} and Occlum~\cite{shen_occlum_2020}.

Lack of in-enclave privilege layers prevents efficient enforcement of trust and isolation. As such, LibOS secrets cannot be robustly isolated and protected from buggy or compromised enclave applications, especially in multi-party enclaves. Software-based isolation mechanisms —MPK‑based ~\cite{vahldiek-oberwagner_erim_nodate,melara_enclavedom_2020}, nested enclaves~\cite{park_nested_2020}, and compiler‑based compartmentalization~\cite{shen_occlum_2020}—mitigate some risks but still demand host cooperation, restrict the programming model, may impose performance overheads, or leave enforcement gaps. Addressing this issue allows new capabilities: faster handling of syscalls and exceptions inside enclaves without software recompilation or enclave transitions; adding timers; easy preemptive scheduling and switching of threads; leading to efficient and richer enclave computing.

Comparatively, CVMs avoid these problems by reusing conventional OS stacks, but attempts to shrink the CVM TCB (unikernels~\cite{duy_incognitos_2025}, micro-VMs~\cite{kuvaiskii_gramine-tdx_2024,zhang2018kylinx}, and hardened minimal images~\cite{community_confidential_nodate}) trade one set of compatibility challenges for another without restoring the small, auditable TCB of enclaves.

Recent CVM extensions, AMD SEV-SNP's Virtual Machine Privilege Levels (VMPLs)~\cite{noauthor_sev_2022} and Intel TDX partitioning~\cite{noauthor_intel-tdx-module-15-td-partitioning-spec-354807002_nodate}, expose primitives that allow a security monitor~(SM)  to create process-level isolation, protected from the guest OS, in a CVM. Prior related efforts—NestedSGX~\cite{wang_road_2025}, vSGX~\cite{zhao_vsgx_2022}, Veil~\cite{ahmad_veil_2023}, and Gramine-TDX~\cite{kuvaiskii_gramine-tdx_2024}—demonstrate important directions but leave the SGX–CVM gap open: they either require modifications of applications  or runtimes, incur large cross-VM transitions, or need hypervisor changes.

\emph{Our insight} is that a single CVM-resident SM can expose the full, unmodified SGX ISA to legacy enclaves and offer new enclave extensions that fix SGX's long-standing limitations---all without hypervisor, hardware, or host-kernel changes. The core idea combines VMPL isolation, a new in-enclave \emph{horizontal privilege separation}, and a kernel driver outside the enclave TCB for resource management.

We present \sys, a CVM firmware built on Coconut SVSM~\cite{noauthor_coconut-svsmsvsm_2025} and AMD SEV-SNP. It provides two enclave modes: (1) bit-for-bit SGX compatibility for unmodified binaries, and (2) a \emph{novel extended SGX} mode with; horizontal privilege levels, copy-on-write cloning, in-enclave exception handling, direct memory management, soft timers, short syscall loops, new context-switch instructions, and SGX-style attestation while keeping resource management under guest OS.

We implement \sys as a Rust SM and Linux kernel driver, evaluated on AMD EPYC 7313. An \texttt{EENTER}/\texttt{ERESUME} takes 9.7~$\mu$s (dominated by a 4.5--5.5~$\mu$s VMPL transition). Compute/memory workloads see 4.8\% average overhead on nbench; no overhead on LMbench \texttt{bw\_mem}, WolfSSL, or SGXGauge hash-join. \sys adds under 5\% overhead on Redis (SCONE), improves Nginx throughput by 25\% (Gramine EXITLESS~\cite{gramine_docs}), and incurs ${\sim}$12\% on MariaDB (SCONE). COW-Cloning scales linearly (93~$\mu$s/MB), allowing elimination of costly page re-encryption steps in \texttt{fork}~\cite{gramine_docs,baumann_fork_2019}.

This paper makes the following contributions:
\begin{enumerate}[noitemsep]
    \item \textbf{Zero-modification SGX on CVMs.} \sys executes unmodified SGX binaries, and runtimes (e.g., Gramine, SCONE) in a CVM, enabling drop-in adoption of the SGX ecosystem.
    \item \textbf{Extended SGX (eSGX) novelties.} eSGX introduces horizontal privilege levels, architectural enclave cloning, in-enclave exception handling and direct memory management, soft timers, and short syscall paths to address core SGX overheads and gaps.
    \item \textbf{Minimal-TCB SM.} \sys excludes the guest OS from the enclave TCB, keeps resource management under the untrusted host, and composes with existing CVM attestation without requiring hypervisor or hardware changes.
    \item \textbf{Implementation and evaluation.} We implement \sys on Coconut SVSM and SEV-SNP and demonstrate low overheads on compute- and memory-bound workloads, modest overhead on I/O-bound applications, and significant speedups for workloads previously dominated by exits, demonstrating practicality for production use.
\end{enumerate}
\section{Background}
\subsection{Intel SGX}
\label{sec:intel_sgx_background}

Intel Software Guard Extensions (SGX) isolates regions of virtual address space, \emph{enclaves}, from the OS and other untrusted code. Enclave memory resides in encrypted Processor Reserved Memory (PRM): the OS maps enclave virtual addresses into the Enclave Page Cache (EPC), and integrity is enforced through Enclave Page Cache Metadata (EPCM). Enclaves are managed via \texttt{ENCLS} (ring~0) and executed via \texttt{ENCLU} (ring~3). Thread Control Structures (TCS) define per-thread execution contexts. At initialization, SGX measures enclave code and data into \texttt{MRENCLAVE} and can emit reports bound to this measurement for local or remote attestation; Quoting Enclaves (QEs) sign reports with secure keys~\cite{costan_intel_2016}.

\textbf{Exception model.} Any exception or interrupt during enclave execution triggers an \emph{asynchronous exit} (AEX): the CPU saves enclave state into the current state save area (CSSA) frame, clears sensitive state, and leaves enclave mode. The OS may handle the event or deliver a signal to a user-space AEX handler. \texttt{ERESUME} restores the saved state. SGX enclave dynamic memory management (EDMM) relies on AEX for resource management with the OS.

\subsection{AMD SEV-SNP}
\label{sec:amd_sev_snp_background}

AMD SEV-SNP, is the latest AMD's CVM extension that builds on SEV/SEV-ES memory and register protections by adding integrity protection, secure interrupt handling, and flexible attestation using an isolated Secure Processor (SP) running firmware independent of normal CPU cores.

\textbf{Reverse Map Table (RMP).} The SP-managed RMP injectively maps every System Physical Address (SPA) to a Guest Physical Address (GPA) while recording ownership and permissions. A page is owned by the hypervisor (HV), a guest VM, or the SP firmware. Writes are allowed only by the owner; HV reads return encrypted data. The \texttt{RMPUPDATE} instruction and the SEV-SNP API modify RMP entries. To prevent remap attacks, every page must be explicitly accepted via \texttt{PVALIDATE} inside the CVM; use of an unvalidated page raises a VMM Communication Exception (\texttt{\#VC}).

\textbf{Virtual Machine Privilege Levels (VMPLs).} This feature provides hierarchical privilege domains in a CVM. It has 4 levels (VMPL0-VMPL3) each with a Virtual Machine Save Area (VMSA). Per-page permissions (read, write, user- and supervisor-executable) are recorded per VMPL in RMP table. VMPL0 is the most privileged level; the CVM boots into VMPL0 with full permissions granted on all memory pages. VMPL0 creates higher VMPLs via \texttt{RMPADJUST} and \texttt{PVALIDATE} on private pages, and initializes VMSAs before updating the RMP. Permissions checks enforce RMP permissions. 

\textbf{Interrupts.} SEV-SNP has two interrupt interfaces. \emph{Restricted injection} with a doorbell exception (\texttt{\#HV}) polling a paravirtualized queue. \emph{Alternate injection} uses guest-controlled queuing and injection, and a VMSA  to emulate APIC/x2APIC.

\textbf{VMPL switching.} \texttt{VMRUN} binds a vCPU to a specific VMPL via a VMSA pointer in the Virtual Machine Control Block (VMCB); only one VMPL is active per vCPU at any time. I/O and \texttt{CPUID} instructions force a non-automatic exit to VMPL0 for handling when REFLECT \#VC attribute is configured. The guest may also request the HV for a VMPL switch via the Guest-Host Communication Block (GHCB).

\textbf{Attestation and sealing.} At \texttt{VMLAUNCH}, the SP measures all guest data and metadata pages into a Launch Digest; an optional signed Identity Block (IDB) can bind expected digests. After launch, the guest obtains attestation reports from the SP over a secure channel, signed by the Versioned Chip Endorsement Key (VCEK) or Versioned Loaded Encryption Key (VLEK) registered with the AMD Key Distribution Service. Sealing keys can be derived from VCEK, VLEK, or a migration-agent-supplied VM Root Key (VMRK).

\subsection{Secure VM Service Module (SVSM)}
\label{sec:svsm}

SVSM defines an execution environment for hosting privileged VMPL0 service modules, and protocols to discover and interact with them. The SVSM binary, VM secrets, and metadata are measured at launch; additional firmware or binaries can be loaded at other VMPLs under the same measurement model. COCONUT-SVSM~\cite{noauthor_coconut-svsmsvsm_2025} is an open-source, Rust-based SVSM reference implementation of the protocol.

\subsection{Library OSes for SGX}
\label{sec:libos_background}

Running unmodified applications in SGX usually requires a LibOS that services syscalls inside the enclave. Representative systems range from libc shims to full in-enclave kernels: Panoply~\cite{shinde_panoply_2017}, SCONE~\cite{scone}, Gramine (formerly Graphene-SGX)~\cite{tsai_graphene-sgx_2017}, SGX-LKL~\cite{priebe_sgx-lkl_2020}, and Occlum~\cite{shen_occlum_2020}. SGXv2~\cite{mckeen2016intel} added EDMM and richer exception information, easing LibOS integration, but the architectural constraints that motivate \sys's design; host-mediated exceptions; lack of intra-enclave privilege and isolation between an enclave application and LibOS, page-table control, and architectural enclave cloning support; and coarse AEX-based preemption, are unaddressed in SGXv2 and are revisited in~\cref{sec:sgx-binary-compatibility}.

\section{Overview}

\sys is designed to execute unmodified SGX binaries transparently inside AMD SEV-SNP CVMs, while introducing new primitives that strengthen SGX's security properties and address its architectural limitations. This section states the design goals, threat model, and overall architecture.

\subsection{Design Goals}
\label{sec:design_goals}

\sys pursues five goals:

\noindent{\cnumber{1} \textbf{Userspace binary and ISA compatibility.}} Unmodified SGX binaries, SDKs, and LibOSes run on \sys without recompilation. \sys implements the Intel SGX ISA and matches its execution semantics.

\noindent{\cnumber{2} \textbf{Memory flexibility.}} \sys lets any guest memory back an enclave, enabling dynamic mapping of guest OS memory as private enclave memory unlike SGX's fixed EPC.

\noindent{\cnumber{3} \textbf{Rich LibOS abstractions.}} \sys  brings new ISA extensions and features for greater LibOS capabilities while retaining SGX security and resource control model.

\noindent{\cnumber{4} \textbf{Strengthened isolation.}} A dedicated VMPL hosts enclaves, hardening the boundary between enclave, SM, and untrusted guest OS.

\noindent{\cnumber{5} \textbf{Attestation support.}} \sys exposes SGX-compatible attestation workflows so applications can verify they run on genuine, attested hardware and CVM. 

\subsection{Threat Model}

\sys minimizes the trusted surface while accounting for adversaries both inside and outside the CVM. \sys adopts SGX and AMD SEV-SNP threat models except hardware attacks outside their scope, and DoS attacks. The TCB includes only the SM and the enclave; the guest OS and its drivers are explicitly outside this boundary. \sys defends against four adversary classes:

\noindent\textbf{Malicious hypervisor.} \sys relies on SEV-SNP hardware primitives to protect enclave memory and page tables from host observation or tampering.

\noindent\textbf{Compromised guest OS.} Even a fully compromised guest kernel cannot access enclave data: the SM enforces isolation through VMPL permissions independently of the guest OS.

\noindent\textbf{Untrusted peer applications.} VMPL hardware isolation protects enclaves from guest OS and other applications.

\begin{figure}[h!]
    \centering
    \includegraphics[width=0.33\textwidth]{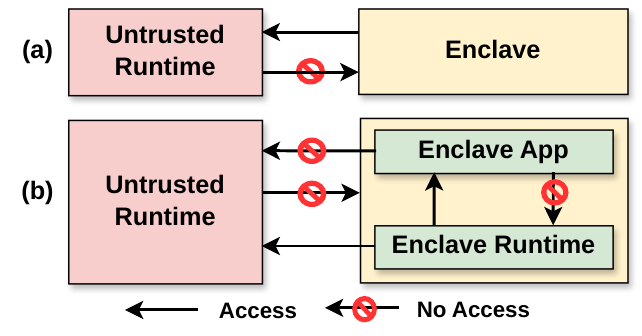}
    \caption{\sys modes: (a) SGX and (b) eSGX split application models showing relative access restrictions of trusted and untrusted parts of an application.}
    \label{fig:teevisor-split-model}
\end{figure}

\noindent\textbf{Trust enforcement of enclave code.} Third-party code running inside an enclave may be malicious, incorrect or buggy. \sys partitions the enclave into a trusted runtime range and an enclave-application user range, enforced by separate page tables so user code cannot corrupt, access or randomly execute in runtime. These ideas are illustrated in \Cref{fig:teevisor-split-model}.
Typically LibOS(e.g., SCONE) spans new \textit{Enclave Runtime} region with the application residing in \textit{Enclave App}.

\noindent{\textbf{Security of eSGX}}. eSGX improves and retains SGX security guarantees by restricting the enclave application (horizontal privilege, switching instructions), from accessing non-enclave memory while VMPL permissions (rather than cache flushing as in SGX) protects access to enclaves for exits. New paths (timers, internal exceptions, syscall) retain SGX SSA-based exception model. Clones have special tokens to ensure correct parent-child matching, while the memory ownership model (see \cref{app:aba}) ensures correct COW enclave-group memory mapping (i.e., no cross-enclave mapping). Rollback protection is guaranteed by platform attestation~(\cref{sec:teevisor-attestation}).

\subsection{Architecture}

\begin{figure}[h!]
    \centering
    \includegraphics[width=0.5\textwidth]{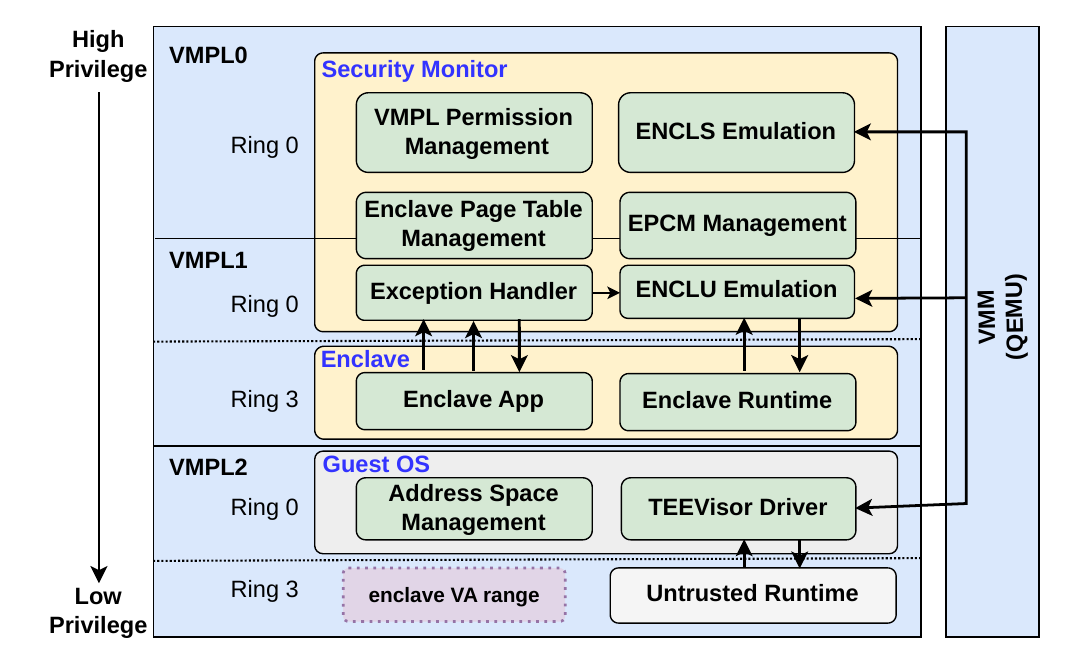}
    \caption{\sys system overview.}
    \label{fig:teevisor-overview}
\end{figure}

\Cref{fig:teevisor-overview} shows \sys's architecture. The SM runs in VMPL0 kernel space and services \texttt{ENCLS} requests from the guest (VMPL2), managing enclaves and granting or revoking VMPL permissions aptly. In VMPL1 kernel space, the SM services \texttt{ENCLU} requests and emulates context switches for enclave transitions; enclave code runs in VMPL1 user space. \texttt{ENCLU} instructions issued from inside an enclave (e.g., \texttt{EACCEPT}) are emulated by the exception handler. With horizontal privilege modes, enclaves can switch between trusted runtime and untrusted user execution with isolated memory. The \sys driver runs in the guest Linux kernel (VMPL2) to trap userspace \texttt{ENCLU} instructions and to expose an \texttt{ENCLS} interface compatible with existing SGX driver. The SVSM protocol is extended to handle complex requests.

\subsection{Comparison with Existing Approaches}
\Cref{tab:state-of-the-art-comparison} compares \sys with three SEV-SNP-based designs. vSGX~\cite{zhao_vsgx_2022} preserves binary compatibility by splitting the application across a normal VM and a CVM; trap handlers in both VMs decode enclave instructions and drive SGX emulation. The cross-VM path introduces large round-trip overheads and requires hypervisor adaptation. Veil~\cite{ahmad_veil_2023} uses VMPLs to create enclaves inside a single CVM with a dedicated VMPL0 monitor, but defines its own SGX-like ABI and relies on hypervisor changes, so it cannot accept unmodified SGX binaries. NestedSGX~\cite{wang_road_2025} extends COCONUT-SVSM to host enclaves in SVSM-reserved memory and emulates SGX via ioctls and syscalls, but it does not fully support the SGX model due to missing binary compatibility and incomplete shared-memory synchronization. \sys also builds on COCONUT-SVSM and dedicates a VMPL to enclaves, but is unique in that it runs unmodified SGX binaries at low VMPL-transition overhead and adds features beyond SGX, including in-enclave exception and syscall handling, and architectural COW cloning with page sharing across enclaves.

\begin{table}[htbp]
    \centering
    \caption{Comparison between \sys and state-of-the-art systems.}
    \resizebox{\columnwidth}{!}{%
        \begin{tabular}{|c|c|c|c|c|c|}
            \hline
            \rowcolor{black} 
            & \makecell{\color{white} Flexible \\ \color{white} memory \\ \color{white} conversion} 
            & \makecell{\color{white} SGX binary \\ \color{white} compatibility} 
            & \makecell{\color{white} No hypervisor \\ \color{white} adaption} 
            & \makecell{\color{white} Low- \\ \color{white} overhead \\ \color{white} switches} 
            & \makecell{\color{white} Internal \\ \color{white} exception \\ \color{white} handling} \\
            \hline
            vSGX~\cite{zhao_vsgx_2022} & \redcross & \greencheck & \redcross & \redcross & \redcross \\
            \hline
            Veil~\cite{ahmad_veil_2023} & \greencheck & \redcross & \redcross & \greencheck & \redcross \\
            \hline
            NestedSGX~\cite{wang_road_2025} & \redcross & \redcross & \greencheck & \greencheck & \redcross \\
            \hline
            \sys & \greencheck & \greencheck & \greencheck & \greencheck & \greencheck \\
            \hline
        \end{tabular}%
    }
    \label{tab:state-of-the-art-comparison}
\end{table}
\section{Design}
We first describe the enclave life cycle (\cref{sec:enclave_life_cycle}) and how \sys achieves userspace binary compatibility~\cnumber{1} and extended SGX (eSGX) features~\cnumber{3} (\cref{sec:sgx-binary-compatibility}). We then present the memory-management design that delivers memory flexibility~\cnumber{2} and strengthened isolation~\cnumber{4} (\cref{sec:memory_management}), and close with \sys's attestation chain~\cnumber{5} (\cref{sec:teevisor-attestation}).

\subsection{Enclave Life Cycle}
\label{sec:enclave_life_cycle}
We describe the control-flow transfers among \sys's parts during enclave creation, execution, and destruction.

\subsubsection{Enclave Creation}
All \texttt{ENCLS} instructions are handled by the SM in VMPL0. Untrusted Runtime interacts with the driver via \textit{ioctls} as in SGX. The driver emulates ENCLS leaf instructions via dedicated SVSM calls. \texttt{ECREATE} creates an enclave in VMPL1 by initializing an SGX Enclave Control Structure (SECS) page, enclave's root identity. SECS specifies the enclave's attributes and enables the Runtime and Clone extensions. \texttt{EADD} adds regular pages with customized permissions or TCS pages to the enclave identified by its SECS. \texttt{EEXTEND} measures each newly added page into the enclave's digest. After all pages are added and measured, \texttt{EINIT} submits a signature structure (SIGSTRUCT) to verify that measurements and attributes match expectations, finalizes the measurement, and makes the enclave executable.

\subsubsection{Enclave Execution}
Enclave execution comprises launch, exit, and destruction.

\begin{figure}[h!]
    \centering
    \includegraphics[width=0.51\textwidth]{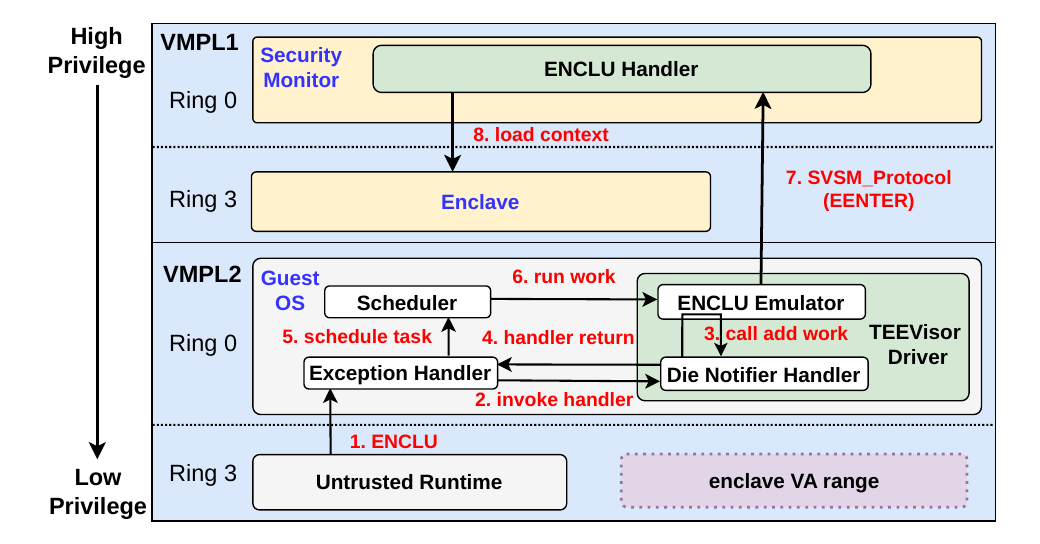}
    \caption{\sys enclave launch control flow.}
    \label{fig:enclave_launch_flow}
\end{figure}

\textbf{Enclave Launch.} \Cref{fig:enclave_launch_flow} shows how \sys preserves binary compatibility during launch. When user-space code executes \texttt{EENTER}, the \sys driver's die-notifier hook, invoked on the \texttt{\#UD} exception path, parses the faulting instruction. If it is an \texttt{ENCLU}, the handler enqueues the emulation task on the current task's work list and marks the exception as handled. Before returning to user space, the kernel executes the emulation, which switches to VMPL1 via the SVSM protocol; the VMPL1 \texttt{ENCLU} handler then loads the enclave context and enters the enclave.

\begin{figure}[h!]
    \centering
    \includegraphics[width=0.48\textwidth]{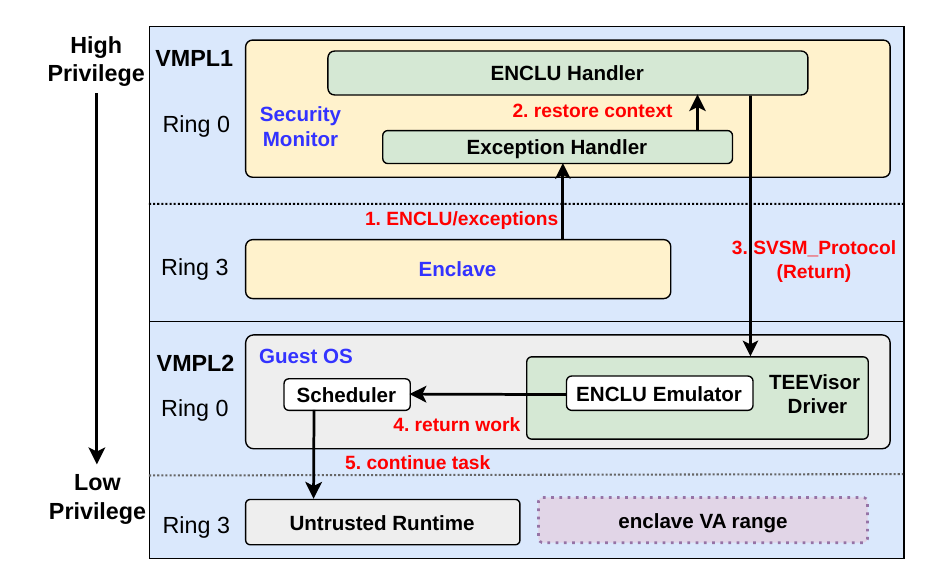}
    \caption{\sys enclave exit control flow.}
    \label{fig:enclave_exit_flow}
\end{figure}

\textbf{Enclave Exit.} A synchronous exit can occur on \texttt{EEXIT} or a new \texttt{ECLONE} (\cref{sec:design_enclave_cloning}). An AEX occurs on exceptions or \texttt{ERAISE}. As in \cref{fig:enclave_exit_flow}, both exits are handled by the VMPL1 SM exception handler, which then returns to VMPL2. The \sys driver's \texttt{ENCLU} emulator then updates the task context from the protocol parameters and prepares execution for resumption. For AEXes, the emulator decodes the exception vector from the protocol parameters and attempts to resolve the fault internally rather than invoking the native Linux handler; if \sys cannot resolve it, the correct signal is delivered to the untrusted task. Finally, the scheduler resumes the untrusted task with updated context.

\subsubsection{Enclave Destruction}
The \sys driver initiates destruction when the guest OS terminates the host process. \texttt{ECCLEARCACHE}, \texttt{ESYNC}, and \texttt{EREMOVE} reclaim cache blocks, shared pages, and other resources from the SM; all remaining state is released once the SECS page is removed.

\subsection{Binary Compatibility and eSGX}
\label{sec:sgx-binary-compatibility}

To meet~\cnumber{1}, \sys preserves the SGX binary interface and emulates SGX semantics precisely: \texttt{ENCLU} instructions in untrusted code are trapped by the \sys driver at VMPL2, while those inside enclaves are trapped by the SM at VMPL1. This uniform trapping mechanism also lets \sys correctly emulate instructions that hardware cannot execute directly. Beyond binary compatibility, \sys delivers eSGX through five features (F1--F5). It is enabled by two new \texttt{SECS} flags; \texttt{Runtime} and \texttt{Clone}. Jointly, F1--F5 address the SGX limitations summarized in~\cref{sec:libos_background} and advance goals~\cnumber{1}--\cnumber{4}:

\noindent\textbf{F1: Horizontal privilege layers.} The enclave address space is split into \emph{Enclave Runtime~(ER)} and \emph{Enclave Userspace~(EU)}. ER can read and modify EU memory, but not vice versa, and holds the privilege to execute enclave-management instructions and handle exceptions that enable running an unmodified application in EU with low overhead.

\noindent\textbf{F2: In-enclave memory management.} ER manages enclave memory attributes inside, reducing exits and speeding applications with frequent attribute modifications.

\noindent\textbf{F3: Flexible exception handling.} Exceptions raised inside the enclave are handled by Runtime handlers when configured; unresolved events can be re-raised to the host. Unconfigured exception handlers trigger a normal AEX.

\noindent\textbf{F4: Controlled runtime/user transitions.} \sys provides syscall- and \texttt{iret}-like instructions enabling: (1) transitions between ER and EU; (2) triggering correct Runtime handler,~and (3) enabling a mode or context switch.

\noindent\textbf{F5: Enclave cloning.} \sys supports page-granularity COW cloning of a running enclave, preserving integrity and preventing misuse by untrusted parties.

\subsubsection{Enclave Runtime}
\label{sec:enclave_runtime}

ER enforces isolation at memory and instruction-privilege levels. Code in ER can access the whole enclave; EU code sees user range only, with the runtime range placed at higher (non-overlapping) addresses. Metadata pages (SSA, TCS, handler pages) and the enclave entry point must reside in the runtime range; execution begins in runtime mode. Shared memory with guest OS is accessible only from the ER. A dedicated \emph{User SSA} (USSA) saves EU context alongside the existing SSA range and is repointable via \texttt{ESETUSSA}. In EU, \texttt{ENCLU} is illegal (\texttt{\#UD}) while \texttt{syscall} is legal and triggers a mode switch. Since \sys owns the enclave page table, \texttt{EMODPE} and \texttt{EMODPR} are replaced with \texttt{EMODP} for permission control without guest OS interaction.

\begin{figure}[h!]
    \centering
    \includegraphics[width=0.48\textwidth]{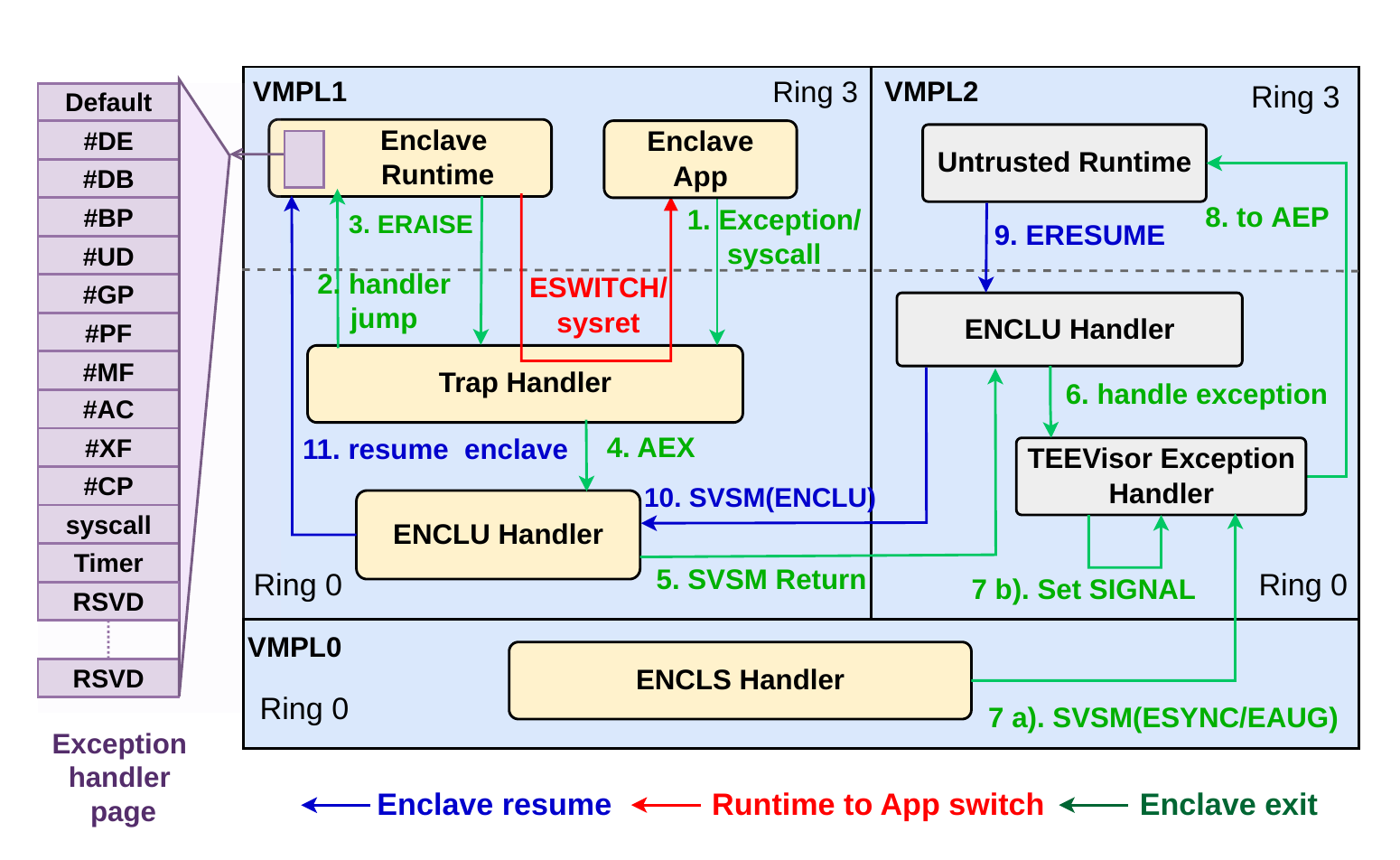}
    \caption{\sys runtime exception handling flow.}
    \label{fig:enclave_exception_handling}
\end{figure}

\textbf{Syscall and exception fast paths.} \cref{fig:enclave_exception_handling} shows in-enclave syscall and exception handling. The trap handler redirects an EU \texttt{syscall} to a syscall entry in the handler page, switching modes; \texttt{sysret} switches back to EU.~SGX-supported exceptions (with vectors in SSA) redirect to their corresponding entries, incrementing CSSA and saving context in the SSA. Other interrupts and guest OS shared-memory \#PFs follow AEX flow. To request guest OS support (e.g., handling \#PF via \texttt{ESYNC}/\texttt{EAUG}), \texttt{ERAISE} re-raises SGX-supported exceptions to trigger an AEX. \texttt{ESWITCH} restores EU context in USSA.

\subsubsection{Enclave Clone (F5).}
\label{sec:design_enclave_cloning}
Clone extends the Runtime feature to accelerate child-enclave creation via COW; initially duplicating only metadata pages. It offers consistency and reliability by enforcing a single active TCS at clone, before the SM freezes enclave state. \texttt{ECLONE} freezes an enclave, exits synchronously with context saved and writes \texttt{ECloneMetadata} to guest OS buffer.

An ioctl binds the child using this metadata, and the \sys driver drives the child's \texttt{ECCREATE}/\texttt{ECLONEINFO}/\texttt{ECADD}/ \texttt{ECINIT} sequence during a \texttt{fork}, unfreezing the parent on success or calling \texttt{ECABORT} on failure. Post-clone, \texttt{ERESUME} continues execution on the cloned TCS. Regular pages are shared until a write or EPCM change triggers a COW, which is serviced from an in-enclave block cache (\texttt{ECADDCACHE}), synchronizing mapping states via a polled slotted page. \Cref{app:clone-driver} details the full cloning sequence (\cref{fig:clone_sequence}).

\subsubsection{SGX Instruction Emulation}

\Cref{tab:supported-instructions} lists every SGX instruction supported in \sys. Because \sys disallows memory swapping in favor of flexible memory conversion (\cref{sec:memory_management}), the swap-related \texttt{ENCLS} instructions are not supported. AEXNOTIFY, and its associated \texttt{EDECCSSA}, is likewise unnecessary for Runtime-enabled enclaves, which observe exceptions directly. Two new common instructions, \texttt{ESYNC} and \texttt{EADDB} (\cref{sec:common_new_instructions}), fill emulation gaps. Some legacy instructions are disabled under the Runtime feature to harden the enclave against a malicious guest OS, since the SM manages page attributes without exiting to the host.

\subsubsection{Common New Instructions}
\label{sec:common_new_instructions}

\sys introduces two new ENCLS instructions common to all enclave modes: \texttt{ESYNC} and \texttt{EADDB}. \texttt{ESYNC} manages shared pages inside the enclave and is callable only by the driver, to preserve userspace compatibility and SGX memory guarantees. \texttt{EADDB} accelerates batch page initialization by using an \texttt{SVSM} buffer to describe contiguous page runs instead of registers (as in \texttt{EADD}). It processes pages sequentially, updating a count, and returns the total pages added to an enclave.

\subsubsection{Modified and New Instructions (summary)}

To support F1--F5, \sys extends a handful of SGX structures (SECS, TCS, GPRSGX) and \texttt{EADD}/\texttt{EENTER}/\texttt{ERESUME} with fields and sanity checks for the runtime range, USSA, and handler page. It adds six new \texttt{ENCLU} instructions for the Runtime feature, \texttt{syscall}/\texttt{sysret} for fast mode transitions, \texttt{ERAISE} for host-assisted exception delivery, \texttt{ESETUSSA} for USSA repointing, \texttt{ESETTIMER} for runtime scheduling, and \texttt{ESWITCH} for mode-aware context restore, plus \texttt{EMODP}, which unifies \texttt{EMODPR}/\texttt{EMODPE} for in-enclave page-permission control. The Clone feature introduces a new SECS attribute bit, requires \texttt{EREMOVE} and \texttt{EMODT} to take an enclave SECS under page sharing, and adds a family of \texttt{EC*} \texttt{ENCLS} instructions, \texttt{ECCREATE}, \texttt{ECADD}, \texttt{ECLONEINFO}, \texttt{ECINIT}, \texttt{ECABORT}, \texttt{ECADDCACHE}, \texttt{ECCLEARCACHE}, \texttt{ECSYNC}, together with \texttt{ECLONE} on the \texttt{ENCLU} side. \Cref{tab:new-instructions} lists the headline new instructions.

\begin{table}[htbp!]
    \centering
    \caption{Headline instructions introduced by \sys. See \cref{app:instructions} for the full catalogue and details.}
    \resizebox{0.85\columnwidth}{!}{%
        \begin{tabular}{|l|l|l|}
            \hline
            \rowcolor{black}
            \textbf{\color{white} Instruction} & \textbf{\color{white} Class} & \textbf{\color{white} Purpose} \\
            \hline
            \texttt{ESYNC}    & \texttt{ENCLS}, common  & Sync shared pages into the enclave \\
            \texttt{EADDB}    & \texttt{ENCLS}, common  & Batch page add at initialization \\
            \texttt{EMODP}    & \texttt{ENCLU}, Runtime & Unified in-enclave permission control \\
            \texttt{ERAISE}   & \texttt{ENCLU}, Runtime & Raise exception to guest OS (AEX) \\
            \texttt{ESWITCH}  & \texttt{ENCLU}, Runtime & Switch runtime/user mode on exit \\
            \texttt{ECLONE}   & \texttt{ENCLU}, Clone   & Begin parent-enclave clone \\
            \hline
        \end{tabular}%
    }
    \label{tab:new-instructions}
\end{table}

\subsection{Memory Management}
\label{sec:memory_management}
TEEVisor separates memory into guest-managed normal memory and SM managed secure memory. VMPL permissions isolate the two regions, while the SM can dynamically convert normal pages into enclave memory on demand. TEEVisor extends EPCM metadata to track page state, sharing, and ownership throughout the enclave lifecycle. During enclave construction, pages are converted and mapped with enclave permissions; during destruction, their mappings and references are released, and pages with no remaining users are cleared and returned to the guest OS. See \Cref{app:memory_management}.
 
\subsection{Attestation}
\label{sec:teevisor-attestation}
\sys uses two attestations: (1) \emph{Platform attestation} to verify CVM setup and identity, provision it, and guard against \emph{rollback attacks}; and (2) \emph{SGX-style attestation} for application enclaves. SM includes embedded metadata to match identity and measurements of a Platform Initialization Enclave~(PIE) and a Quoting Enclave~(QE).

At boot, the SM generates an ephemeral attestation key pair (SMpub/SMpriv) to anchor local and remote attestations. After registering SMpub with a Key Broker Service~(KBS) via SEV-SNP report, SMpriv signs enclave quotes and derives \emph{local attestation} and \emph{ephemeral sealing}~(per boot) keys. KBS's public key (KBSpub), embedded in the boot image, validates the CVM. Currently, \sys \emph{does not support migration}.

\subsubsection{Platform Attestation}
A platform attestation and initialization service~(PAIS) runs on the guest OS to attest the platform via PIE. The SM gives PIE extended privileged access.
PIE fetches a public key and nonce from KBS, requests a VMPL0 report containing hash(SMpub, nonce, MRENCLAVE) as REPORT\_DATA, and submits them back. On verification, SMpub is registered and KBS returns secrets encrypted with it which the PIE provisions the SM with \texttt{ESMPROV} instruction. The SM validates signatures using KBSpub, confirms correct KBS ownership, and accepts the secrets. All subsequent reports signed by SMpriv are accepted by KBS. \Cref{fig:teevisor_platform_attestation} illustrates this flow.

\subsubsection{Local and Remote Attestation}
\sys preserves SGX report structure with compatible local and remote attestation anchored, which use QE, in SEV-SNP hardware and registered keys (SMpub/SMpriv). \Cref{app:attestation} provides the chain-of-trust and key-derivation details.
\section{Implementation}

We implement \sys on top of the COCONUT-SVSM software stack~\cite{noauthor_coconut-svsmsvsm_2025}. In addition to a new Linux kernel driver (\sys Driver, ~3000 LOC) and an extension of the security monitor (\sys Security Monitor, ~11,500 LOC), \sys reuses several VMPL-enlightened components from COCONUT-SVSM to realize its functionality; Linux Kernel, QEMU and Open Virtual Machine Firmware (OVMF).





\subsection{Security Monitor}
\label{sec:security_monitor_implementation}

The SM runs in VMPL0 and services requests from lower VMPLs. It extends the COCONUT-SVSM SM (Rust with supporting assembly). As noted in~\cref{sec:svsm}, SVSM passes parameters through registers plus a registered buffer (CAA page); we add a new request category for extended enclave operations alongside the existing Core, Attestation, and vTPM categories. This subsection covers the implementation details and non-trivial challenges in the SM.

\subsubsection{Security Monitor Initialization Across VMPLs}
\begin{figure}[h!]
    \centering
    \includegraphics[width=0.42\textwidth]{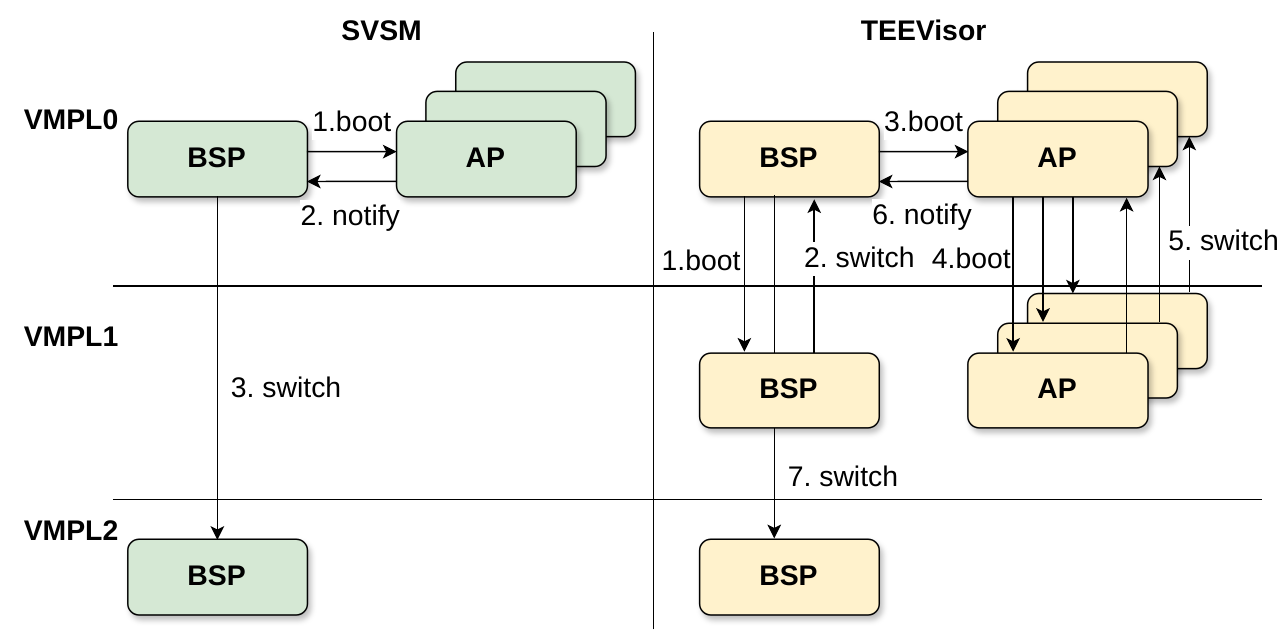}
    \caption{\sys vCPU bootstrap flow.}
    \label{fig:teevisor_bootstrap_flow}
\end{figure}

The SVSM SM runs exclusively in VMPL0 with full VMPL permissions; in \sys it emulates \texttt{ENCLU} in VMPL1 while handling other requests in VMPL0, which demands careful permission setup across both levels (\cref{fig:teevisor_bootstrap_flow}). Initialisation proceeds in two stages. During BSP bring-up, \sys parses the IGVM launch info, grants VMPL1 permissions over the SM range, and installs a VMPL1-specific IDT; each AP then follows the same flow and boots into VMPL1 before returning to VMPL0. Each vCPU keeps VMPL-exclusive per-CPU state that points at a shared CAA page. Once every vCPU is up, the monitor constructs the EPCM from the guest memory layout; its size scales with guest memory, so a sufficiently large heap is reserved (optionally capped).

\subsubsection{Protocol Buffers and ABI}
As \cref{sec:memory_management} explains, VMPL permissions change in an enclave's life cycle. \texttt{ENCLU} requests are handled in VMPL1, thus the SVSM protocol's CAA page must be accessible there; we update VMPL0/VMPL1 mappings and CAA permissions whenever OVMF or the guest OS registers or modifies it.

The SVSM protocol passes up to three register arguments (return value in RAX) and uses pointers for longer structures. For VMPL1 \texttt{ENCLU} requests, VMPL permissions block arbitrary guest page access, restricting CPU state to the CAA page's reserved region and requiring register pointers to match the correct CAA offset. For VMPL0 \texttt{ENCLS} requests, buffer addresses are validated against the \texttt{PVALIDATE} bit (\cref{app:epcm}) to avoid faults on unaccepted pages.

The driver passes physical addresses when emulating across VMPLs because the guest OS lacks enclave page mappings. However, for an \texttt{ENCLU} originating inside the enclave, binary compatibility demands virtual addresses, which the SM resolves by walking its own enclave page tables. The SM returns an SGX error code or an x86 exception vector (encoded) in RAX on return to the guest OS.

\subsubsection{Exception Model}
\label{impl:exception-model}
Logically a single vCPU switches between VMPLs, but the HV views the vCPU's VMPLs as multiple vCPUs with separate contexts that share one APIC ID. Our baseline SVSM lacks Alternate Injection, so \sys uses Restricted Injection: KVM emulates the APIC and injects interrupts into the target VMPL via \texttt{\#HV} vector. I/O interrupts always go to the active lowest-privilege VMPL (VMPL2 in the normal path), IPIs are routed to the target vCPU in the same VMPL, and timer interrupts are injected into the originating vCPU. Because guest-OS-armed timers cannot cross VMPL boundaries, \sys arms a VMPL1 periodic timer at 100~Hz to preempt enclave execution into SM, then do an AEX; a one-shot timer would be more faithful but costs ${\sim}$10{,}000 cycles per x2APIC MSR access.

\subsubsection{Page Ownership with Clone Feature}

Under Clone, an EPCM entry can be shared across enclaves, and SGX's assumption that every enclave page is removed before its SECS no longer holds. This breaks SECS-based ownership: reusing a destroyed parent's SECS page for a new enclave creates an ABA-style confusion in which shared pages appear to belong to the wrong enclave. \sys resolves this by tracking each enclave via a unique identifier and an active descendant counter in an enclave-node list, which safely validates ownership on every EPCM modification (detailed in \Cref{app:aba}).

\subsubsection{Page Table Optimization}
For eSGX enclaves, the SM keeps two page tables to enforce memory isolation. Because the user range appears in both tables with identical attributes, \sys shares the corresponding page-table entries across the two tables after aligning each level. A reserved bit marks entries that also belong to the runtime table so they are correctly reclaimed when the enclave is removed.

\subsection{\sys Driver}
The driver runs in the Linux kernel at VMPL2 and is invoked by userspace programs to manage enclave life cycles. We preserve binary compatibility and minimize porting effort by retaining the device node \textit{/dev/sgx\_enclave} and its ioctls, and adding new ones for the Clone feature. To run on stock kernels, \sys driver re-implements unexported kernel helpers. Driver implementation details are covered hereafter.

\subsubsection{\texttt{ENCLU} Emulation}

\texttt{ENCLU} instruction raises \texttt{\#UD} exception, which the kernel dispatches through its exception chain. The driver registers a die notifier that hooks every \texttt{\#UD} and, on matching an \texttt{ENCLU}, starts instruction emulation and marks the exception handled. Per Linux conventions the notifier stays short and lock-free, so \sys defers the actual work by enqueueing an emulation task at the head of the current task's work list to run in a safe context. The emulator stores the user context and timer state into the CAA page, switches to VMPL1, and on return reloads the user context and dispatches on the exit reason filled in by the SM: \textbf{Interrupt} (emulator handles the vector before returning to user space so signals are delivered correctly), \textbf{\texttt{EEXIT}} and \textbf{Timer} (returns to user space directly as in  \cref{impl:exception-model}), \textbf{Clone} (\texttt{ECLONE} triggers a synchronous exit and the driver marks the enclave as in-clone), \textbf{Cache} (COW ran out of cache pages, driver calls \texttt{ECADDCACHE}), and \textbf{\texttt{NO\_FREE\_SLOT}} (no free slot on the synchronisation page, driver drains and resumes).

\subsubsection{Shared Memory Synchronization}

\texttt{ESYNC} lazily synchronises shared memory (\cref{sec:common_new_instructions}): the driver tracks shared-memory mappings like enclave pages and unmaps them before destruction. On a shared-memory page fault, the \texttt{ENCLU} emulator's post-AEX page-fault path calls \texttt{handle\_mm\_fault} to populate the mapping and syncs it into the enclave via \texttt{ESYNC}. \sys installs an \texttt{mmu\_notifier} on the enclave's address space; its \texttt{invalidate\_range\_start} hook removes affected shared mappings via \texttt{ESYNC} before the kernel mutates them, and they are re-mapped lazily on next access.

\subsubsection{Clone/Fork Support}
\label{sec:implementation_fork}

During enclave cloning (as shown in \cref{fig:clone_sequence}), the \sys driver multiplexes enclaves on a single fd using an address-space-keyed xarray so parent and child processes can resolve their instances after \texttt{fork}. The workflow is triggered within \texttt{fork} by the VMA's \texttt{vma\_open} callback; if \texttt{CLONE\_BIND} metadata is present, it runs the creation sequence (\texttt{ECCREATE}, \texttt{ECLONEINFO}, \texttt{ECADD}, \texttt{ECINIT}). Failures trigger \texttt{ECABORT}, with the status retrieved via \texttt{CLONE\_RESULT} since \texttt{vma\_open} cannot propagate errors.

COW-Clone copies only private metadata pages. Shared pages are marked read-only with incremented reference counts; subsequent page faults check EPCM permissions to enter the COW path. Shared pages use refcounted structures, cache blocks are freed when relevant pages are released, and a kernel thread synchronizes new COW pages to the guest OS. \Cref{app:clone-driver} details the xarray, ioctls, and state machines.

\subsubsection{Timer}

\texttt{ESETTIMER} avails and runs a per-TCS timer (100$\mu$s resolution) that pauses on OS yields, \texttt{AEX}, or \texttt{EEXIT}, and raises a \texttt{Timer} exception upon expiry.

\section{Evaluation}
\label{experiment_setup}

We evaluated \sys on an AMD EPYC 7313 server (16~physical cores, 2~threads per core, 128~GB DDR4 RAM), with the CVM configured at 16~vCPUs pinned to 3~GHz and 8~GB of RAM. Host and guest ran Ubuntu 22.04 with Linux 6.11.0 (SVSM branch, commit \texttt{3b04c21}); the \sys SM was built in Rust release mode.  We used Intel SGX SDK v2.27, Gramine v1.9 and SCONE v6.0.4 for workload evaluation. We report micro-benchmarks (\cref{micro_benchmarks}) and real-world applications under Gramine (\cref{real_applications_gramine}) and SCONE (\cref{real_applications_scone}).

\subsection{Micro-benchmarks}
\label{micro_benchmarks}
We first measured the execution time of individual emulated SGX leaf instructions and the overhead of each stage of control-flow transfers across the enclave boundary, and verified correctness with targeted unit tests.

We evaluated \sys on a suite of micro-benchmarks covering diverse workloads : \emph{(i)}~Linux/Unix nbench~\cite{linux_unix_nbench} as a compute- and memory-intensive CPU benchmark; \emph{(ii)}~LMbench~\cite{lmbench} for low-level memory latency and bandwidth; \emph{(iii)}~WolfSSL~\cite{wolfssl_repo} for compute-intensive cryptography workloads; \emph{(iv)}~SGXGauge hash-join benchmark for memory-intensive database workloads; and \emph{(v)}~UnixBench \texttt{fstime} (small-block I/O) and IOzone (large-block I/O) for IO performance.

\renewcommand{\arraystretch}{1.2}

\begin{table}[htbp]
    \centering
    \caption{\sys latency for common instructions.}
    \label{tab:normal-instructions-latency}

    \resizebox{0.85\columnwidth}{!}{%
        \begin{tabular}{|c|l|l|l|}
            \hline
            \rowcolor{black}
            \textbf{\color{white} Group} &
            \textbf{\color{white} Instruction} &
            \textbf{\color{white} \sys} &
            \textbf{\color{white} NestedSGX} \\
            \hline

            \multirow{10}{*}{ENCLS}
            & EADDB (EXTEND) & 17.7 $\mu$s + 25.3 $\mu$s/page & / \\
            & EADDB (NO EXTEND) & 16.5 $\mu$s + 1.0 $\mu$s/page & / \\
            & ECREATE & 22.3 $\mu$s & 8.4 $\mu$s \\
            & EDBGRD & 13.3 $\mu$s & / \\
            & EDBGWR & 12.5 $\mu$s & / \\
            & EINIT & 43.0 $\mu$s & 46 $\mu$s \\
            & EMODPR & 13.4 $\mu$s & 9.9 $\mu$s \\
            & EMODT & 14.7 $\mu$s & / \\
            & EREMOVE & 13.9 $\mu$s & 7.8 $\mu$s \\
            & ESYNC & 15.3 $\mu$s & / \\
            \hline

            \multirow{6}{*}{ENCLU}
            & EACCEPT & 1.7 $\mu$s & / \\
            & EENTER / ERESUME & 9.7 $\mu$s & 5--6 $\mu$s$^{*}$ \\
            & EEXIT & 9.9 $\mu$s & 5--6 $\mu$s$^{*}$ \\
            & EGETKEY & 10.8 $\mu$s & 17 $\mu$s \\
            & EMODPE & 4.0 $\mu$s & / \\
            & EREPORT & 21.5 $\mu$s & 30 $\mu$s \\
            \hline

            \multicolumn{4}{|l|}{\footnotesize $^{*}$ NestedSGX performs an OCALL/ECALL in 10--12 $\mu$s.} \\
            \hline
        \end{tabular}%
    }
\end{table}

\renewcommand{\arraystretch}{1.2}

\begin{figure*}[htbp]
    \centering
    \begin{minipage}[b]{0.5\textwidth}
        \centering
        \includegraphics[width=\linewidth]{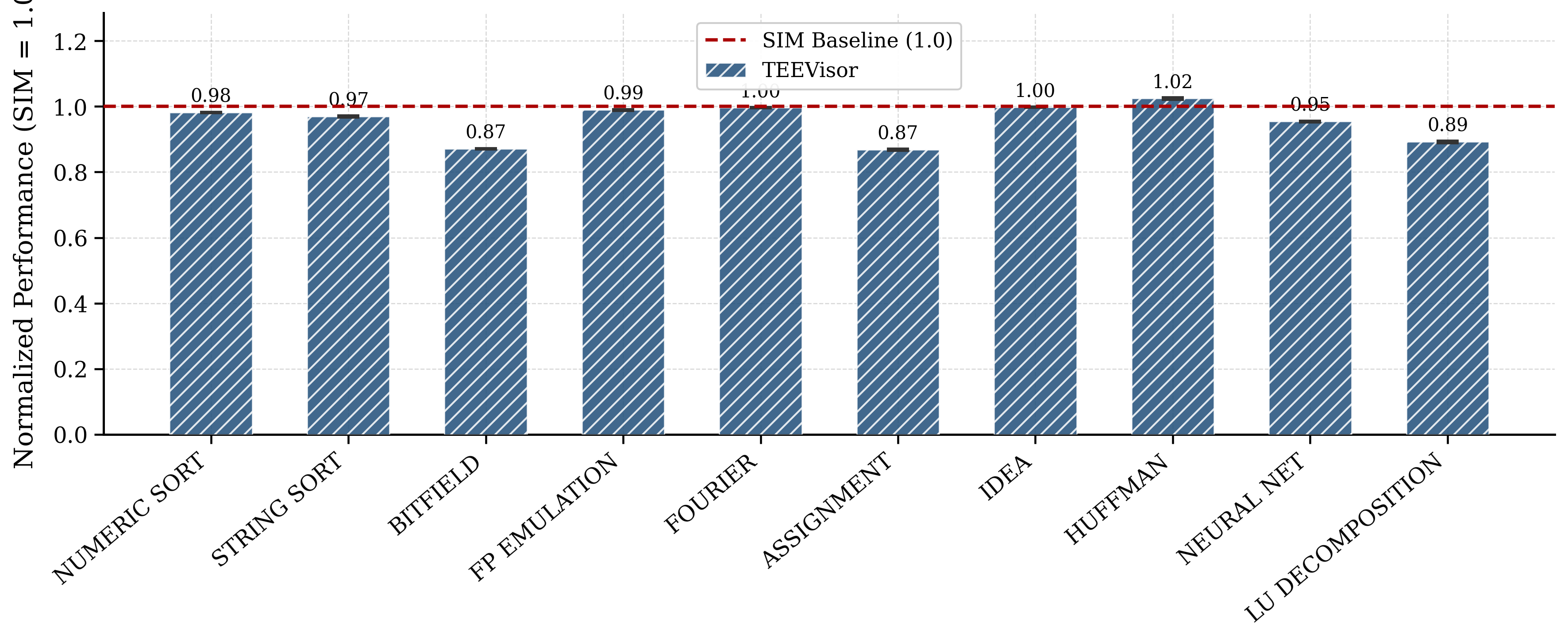}
        \subcaption{nbench normalized to SIM.}\label{fig:nbench}
    \end{minipage}\hfill
    \begin{minipage}[b]{0.5\textwidth}
        \centering
        \includegraphics[width=\linewidth]{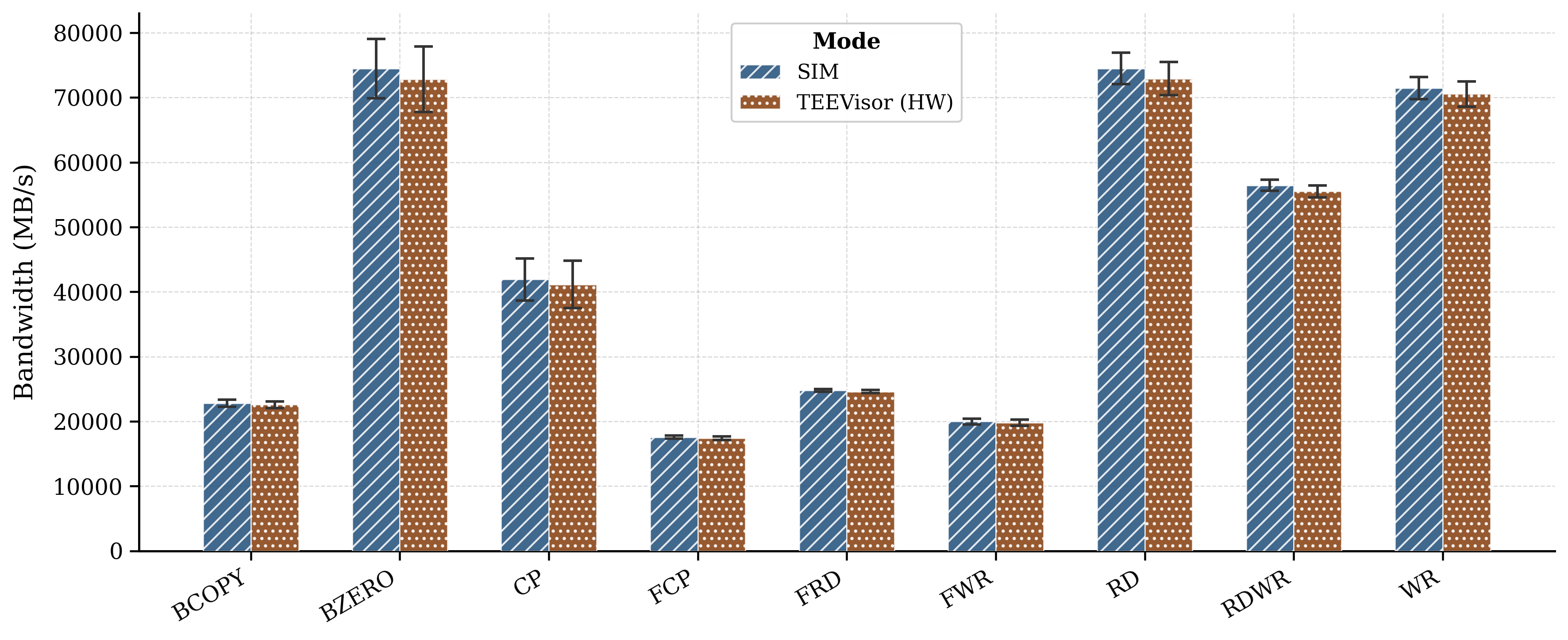}
        \subcaption{LMbench \texttt{bw\_mem} bandwidth.}\label{fig:bw_mem}
    \end{minipage}\hfill
    \begin{minipage}[b]{0.68\textwidth}
        \centering
        \includegraphics[width=\linewidth]{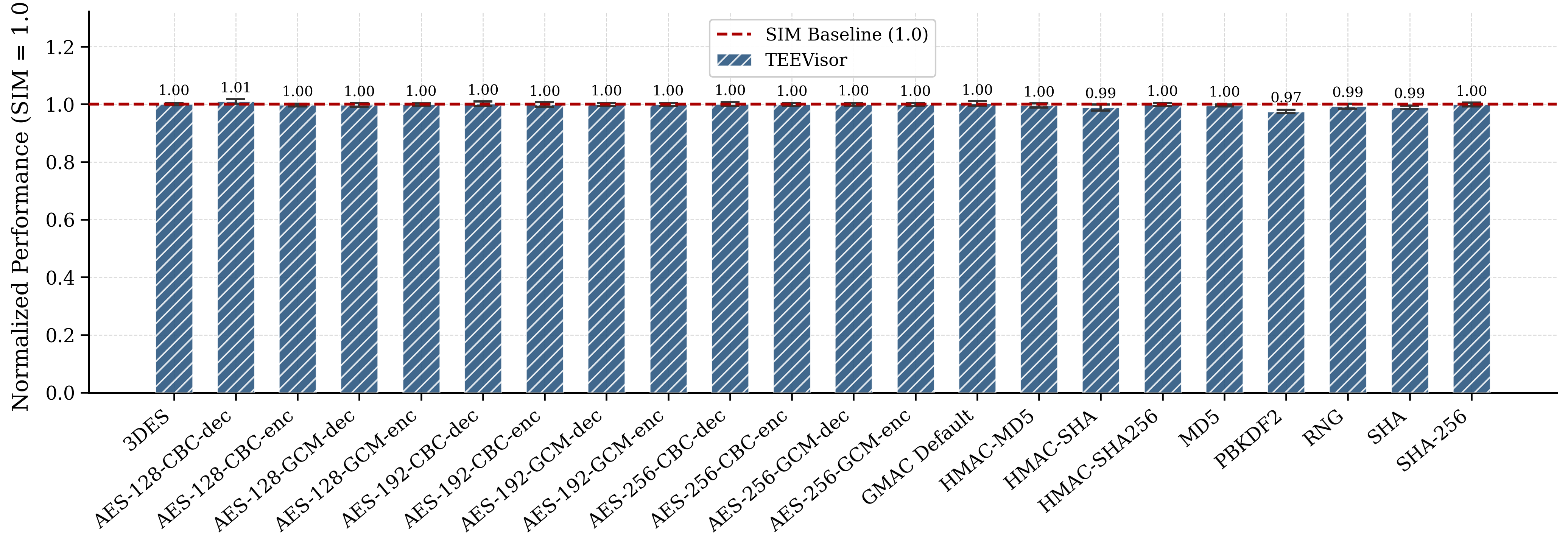}
        \subcaption{WolfSSL normalized to SIM.}\label{fig:wolfssl}
    \end{minipage}
    \caption{Compute- and memory-intensive microbenchmarks.}
    \label{fig:compute_microbench}
\end{figure*}
\vspace{-1pt}
\subsubsection{Instruction Performance} ~~~~~~~~~~~~~~~~~~\\
\textbf{Normal Instructions.}
\autoref{tab:normal-instructions-latency} reports the latency of the instructions available in both SGX and eSGX. For \texttt{ENCLS}, most of the latency came from the two VMPL transitions. \texttt{EADDB} combines \texttt{EADD} and \texttt{EEXTEND} and therefore scales with the number of pages added. \texttt{EINIT}, \texttt{EREPORT}, and the measurement phase of \texttt{EADDB} incurred higher latency due to their cryptographic work.
\sys exhibits higher latency than NestedSGX for ENCLS instructions, primarily due to (1) a higher VMPL switch overhead (5~$\mu$s vs.\ 3~$\mu$s), potentially caused by the use of an additional VMPL; (2) extra driver management and emulation code with sanity checks required for binary compatibility; and (3) additional SM logic needed to support eSGX compatibility. The modification of userspace code in NestedSGX eliminates the compute in the driver needed to distinguish instructions from each other.

\textbf{Runtime and Clone Instructions}.
Runtime instructions in VMPL1 only (\texttt{EMODP}~4.4~$\mu$s, \texttt{ESETUSSA}~3.4~$\mu$s, \texttt{ESWITCH}~4.1~$\mu$s, in-enclave \textsc{exception}~2.5~$\mu$s, \texttt{syscall}~2.3~$\mu$s, \texttt{sysret}~1.1~$\mu$s) avoid boundary crossings and are thus fast; \texttt{ERAISE} is the exception at 7.8~$\mu$s because it exits to the host. \texttt{syscall} and \texttt{sysret} are emulated via \texttt{\#UD} trapping, hence slower than a native syscall. Enclave cloning scales linearly at rate of 93~$\mu$s/MB of enclave size (for a 4-1024~MB user + 64~MB runtime) measured at the driver hook, ~20x faster than enclave creation, dominated by driver management code and \texttt{ECINIT} (114.8~$\mu$s + 8.1~$\mu$s per ~MB of page-table content). \texttt{ECADD} and \texttt{ECLONE} scale with metadata-page count (${\sim}$1.1 and 0.1~$\mu$s/page). \texttt{ECADDCACHE}/\texttt{ECCLEARCACHE} (343.8 / 386.4~$\mu$s, each updating EPCM and RMP VMPL permissions for every page in a 4~MB block) are triggered later in COW handling. Full per-instruction breakdowns are in \cref{app:latency-breakdown} and \cref{app:unit-tests}.

\textbf{Control Handover Stages.}
The \sys driver and the SM each take roughly 2--3~$\mu$s per traversal, while the VMPL1--VMPL2 transition takes 4.5--5.5~$\mu$s and dominates boundary-crossing overhead. AEX emulation in the driver varies with exception type and is therefore not included. The full per-stage breakdown (for \texttt{EENTER}/\texttt{ERESUME}, \texttt{EEXIT}, and AEX) is in \cref{tab:measurement_stages} (\cref{app:latency-breakdown}).

\begin{figure}[htbp]
    \centering
    \begin{minipage}[b]{0.5\columnwidth}
        \centering
    	\includegraphics[width=\linewidth]{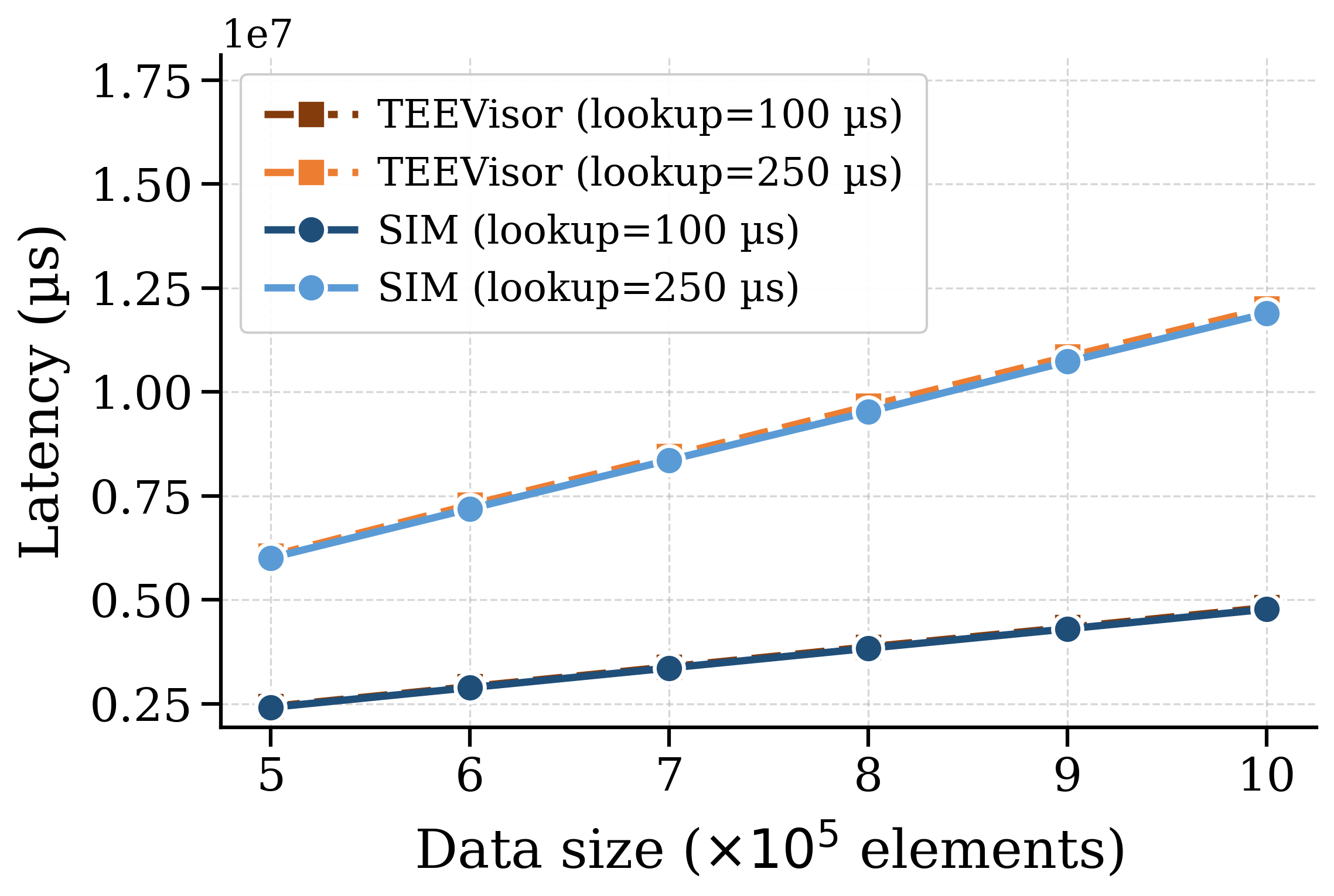}
     	\subcaption{Hash-join lookup latency.}\label{fig:hash_join}
    \end{minipage}\hfill
    \begin{minipage}[b]{0.5\columnwidth}
        \centering
        \includegraphics[width=\linewidth]{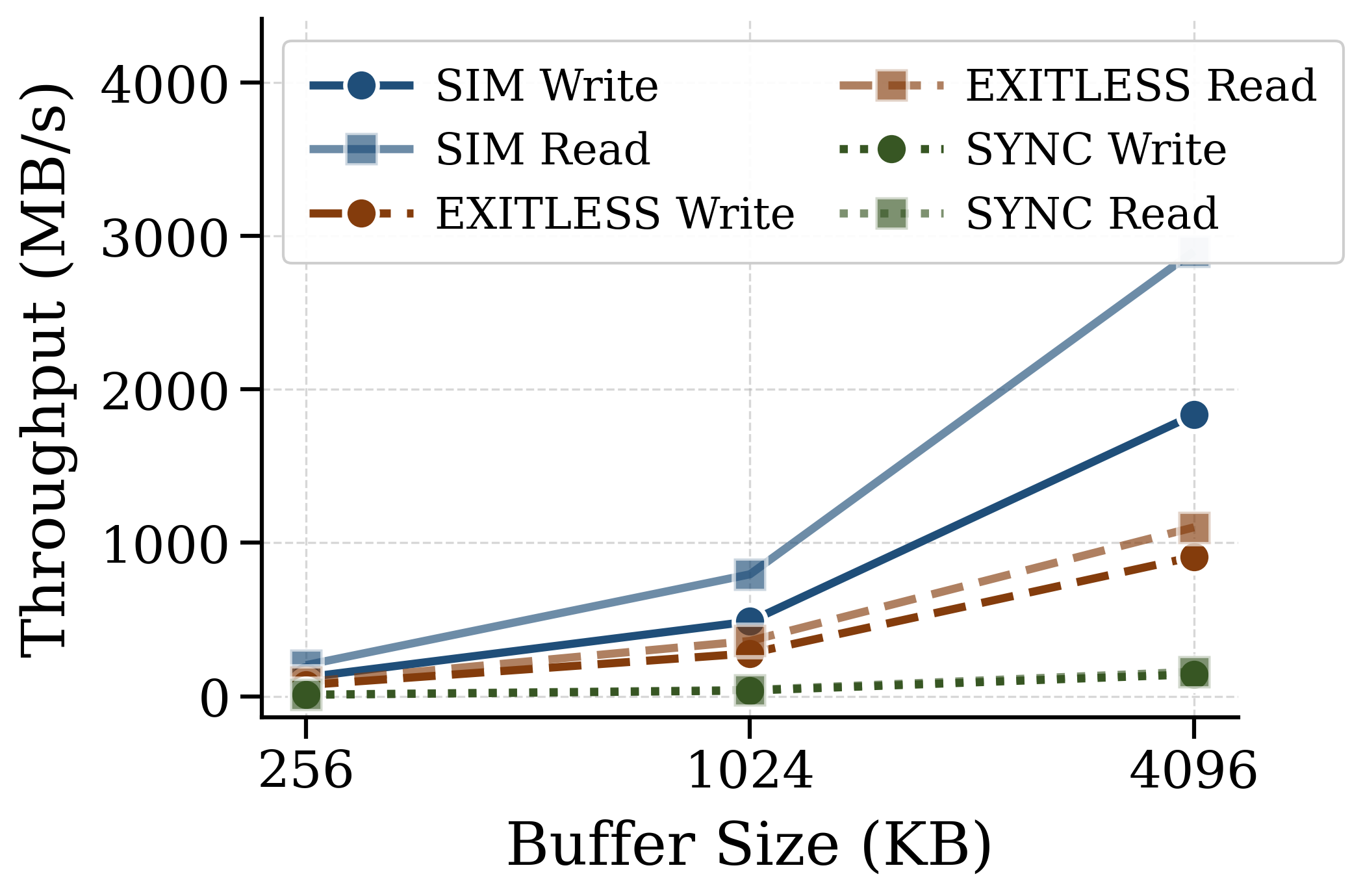}
        \subcaption{UnixBench \texttt{fstime}.}\label{fig:fstime}
    \end{minipage}\hfill

    \caption{Hash-join lookup and UnixBench performance.}
    \label{fig:fs_io}
\end{figure}

\addvspace{-3pt}
\subsubsection{Correctness Tests}
We verified correctness of every native-SGX path and every new control-flow path introduced by \sys with targeted unit tests: vDSO self-tests for native SGX, plus instruction-level positive and negative tests for each Runtime and Clone instruction (\cref{app:unit-tests}). All tests pass.

\subsubsection{Clone and COW Performance}
We provide a detailed evaluation of enclave cloning and COW performance in ~\cref{app:clone_performance}. We show that enclave cloning scales linearly with enclave size at approximately 93~$\mu$s/MB, while clone+COW remains about 1.4--1.7$\times$ faster than creating a new enclave for the evaluated enclave sizes in addition to memory savings it provides.

\subsubsection{Compute and Cryptography Microbenchmarks}
We used three compute- and memory-intensive microbenchmarks to measure overheads outside the syscall path. Linux/Unix nbench~\cite{linux_unix_nbench}, run via sgx-nbench~~~\cite{sgx_nbench}, exercises CPU, FPU, and memory subsystems; \sys adds 4.8\% average overhead over the SIM baseline, primarily dominated by context-switching cost (\cref{fig:nbench}). LMbench \texttt{bw\_mem}~\cite{lmbench} via sgx-bench~\cite{sgx_bench} with a 2~MB block size shows no measurable memory-bandwidth overhead (\cref{fig:bw_mem}); the small differences come from extra OCALLs used for in-enclave time measurement. WolfSSL's built-in cryptographic benchmarks~\cite{wolfssl_repo} match SIM throughput (\cref{fig:wolfssl}).

\begin{figure}[htbp]
	\centering
	\includegraphics[width=.45\textwidth]{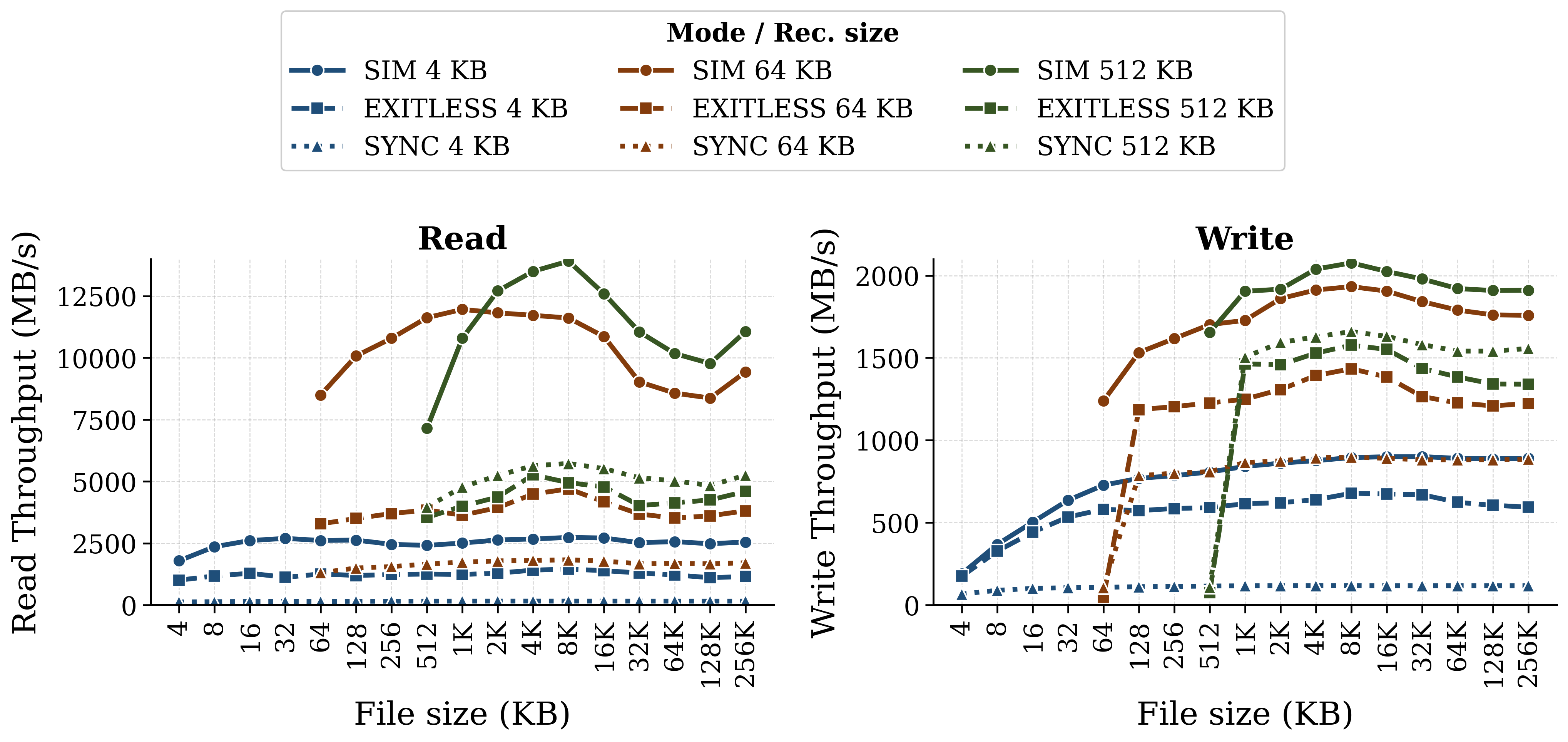}
	\caption{IOzone Throughput.}\label{fig:iozone}
\end{figure}

\subsubsection{SGXGauge Hash Join}
SGXGauge~\cite{sgxgauge} is a benchmark suite for Intel SGX. We used its native hash-join benchmark to stress pointer-chasing memory access: the inner table was built by hashing every element, and probes then hashed keys from the outer table. We fixed the inner table at 100~K entries and varied the outer table from 500~K to 1~M entries, running two rounds that repeated the outer-table lookups 100 and 250 times respectively. As~\autoref{fig:hash_join} shows, \sys adds almost no latency.

\subsubsection{File-System I/O: UnixBench fstime and IOzone}
We measured small- and large-block file I/O under Gramine using UnixBench \texttt{fstime}~\cite{unixbench} and IOzone~\cite{iozone} (4, 64, and 512~KB buffers; 4~KB--256~MB files). As~\cref{fig:fstime} shows, Gramine SYNC suffers from frequent enclave exits (IOzone, 10 runs: EENTER/EEXIT 564,543/564,537 vs. 65.8/59.8 for EXITLESS; AEX 3,002/3,315). EXITLESS incurs buffer-copy and shared-queue overhead, while IOzone (Fig.~\ref{fig:iozone}) shows exitless OCALLs falling back to blocking on timeout, making EXITLESS slower than SYNC at large buffers. Enclave-exit costs diminish as file and buffer sizes grow.

\begin{figure*}[htbp]
    \centering
    \begin{minipage}[b]{0.45\textwidth}
        \centering
        \includegraphics[width=\linewidth]{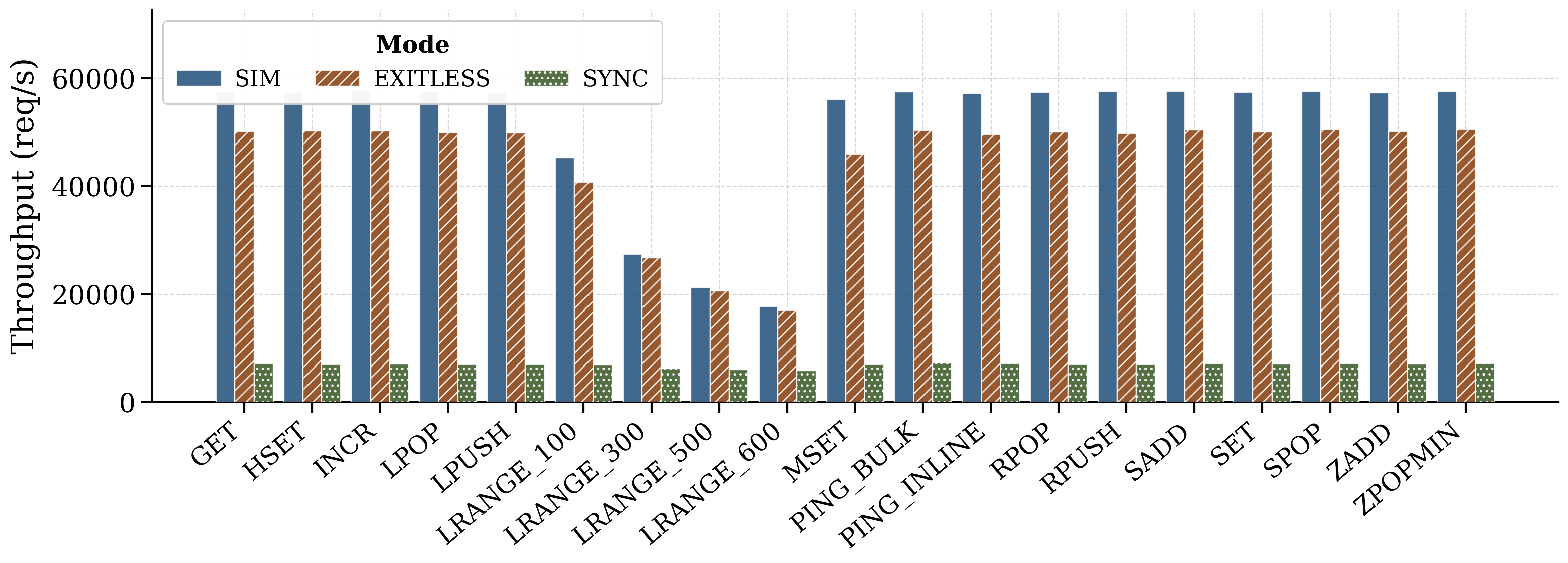}
        \subcaption{Redis.}\label{fig:redis}
    \end{minipage}\hfill
    \begin{minipage}[b]{0.25\textwidth}
        \centering
        \includegraphics[width=\linewidth]{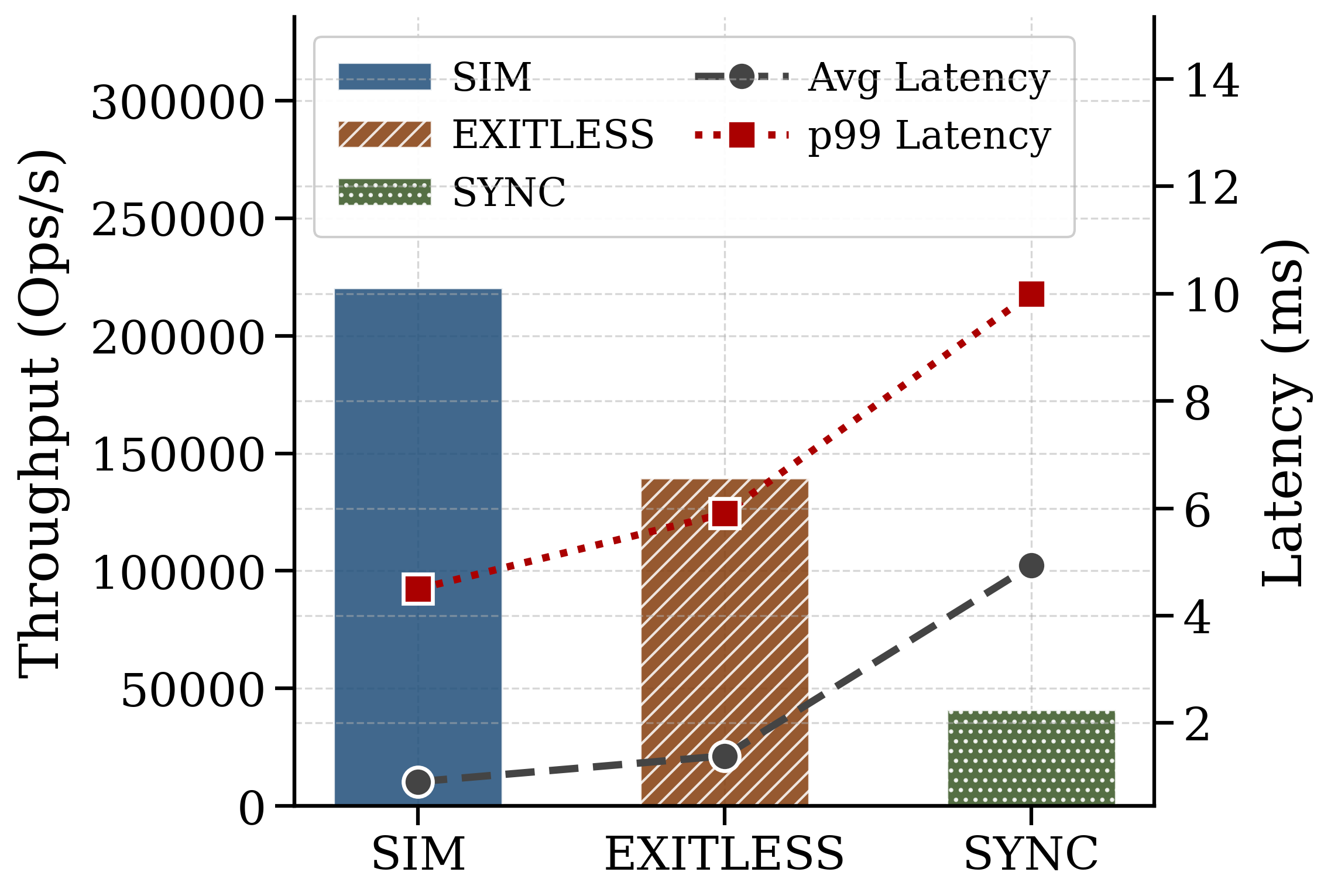}
        \subcaption{Memcached.}\label{fig:memcached}
    \end{minipage}\hfill
    \begin{minipage}[b]{0.26\textwidth}
        \centering
        \includegraphics[width=\linewidth]{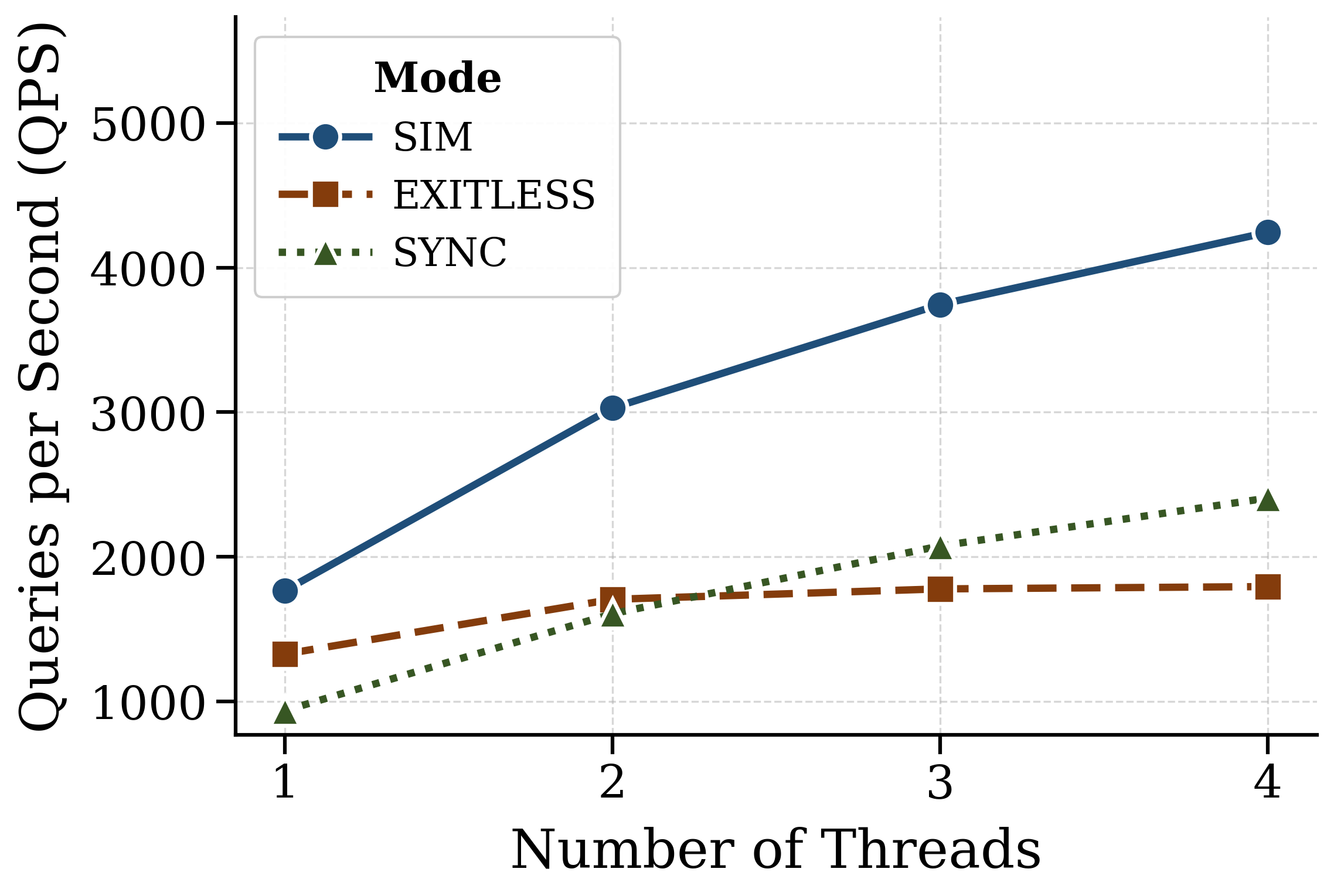}
        \subcaption{MariaDB.}\label{fig:mariadb}
    \end{minipage}
    \caption{Application throughput under Gramine.}
    \label{fig:gramine_apps}
\end{figure*}

\begin{figure*}[htbp]
    \centering
    \begin{minipage}[b]{0.45\textwidth}
        \centering
        \includegraphics[width=\linewidth]{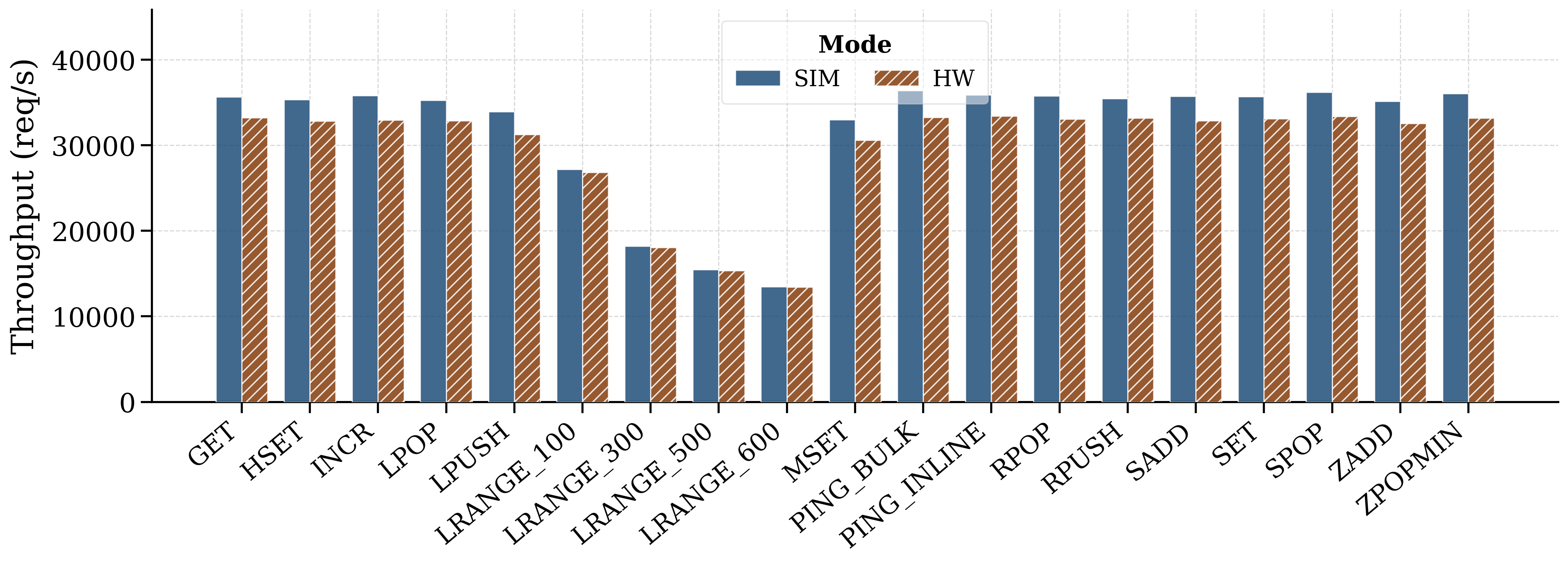}
        \subcaption{Redis throughput.}\label{fig:redis_scone}
    \end{minipage}\hfill
    \begin{minipage}[b]{0.25\textwidth}
        \centering
        \includegraphics[width=\linewidth]{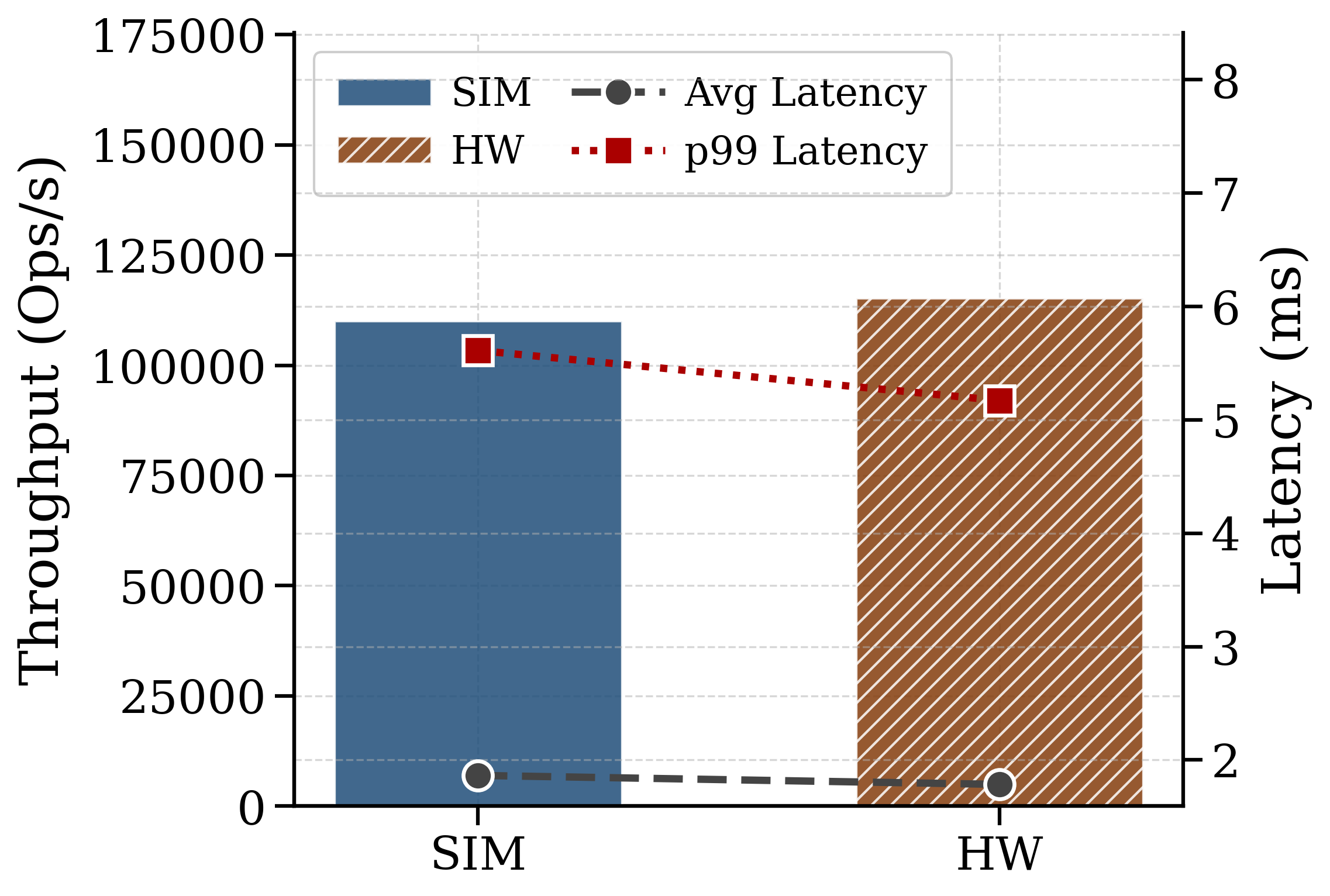}
        \subcaption{Memcached throughput.}\label{fig:memcached_scone}
    \end{minipage}\hfill
    \begin{minipage}[b]{0.26\textwidth}
        \centering
        \includegraphics[width=\linewidth]{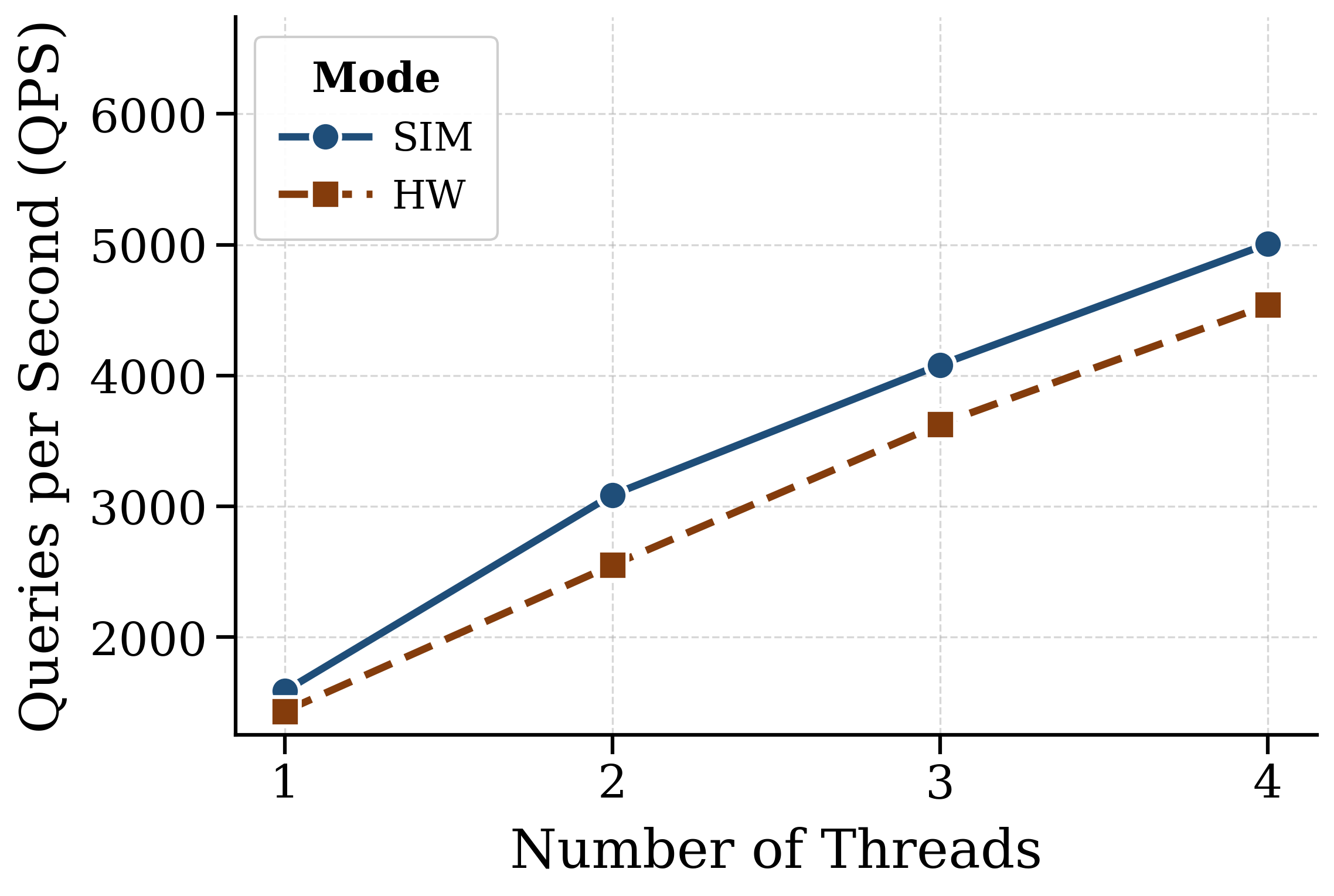}
        \subcaption{MariaDB throughput.}\label{fig:mariadb_scone}
    \end{minipage}
    \caption{Application throughput under SCONE (HW represents TEEVisor).}
    \label{fig:scone_apps}
\end{figure*}

\subsection{Application Benchmarks with Gramine}\label{real_applications_gramine}
We evaluated \sys with Gramine on representative applications: Nginx for high-concurrency HTTP serving and OpenSSL for TLS-heavy cryptography; Redis and Memcached for low-latency, memory-bound key-value stores; and MariaDB for I/O-heavy relational workloads.

Some results are grouped in~\cref{fig:gramine_apps}; we omit \textbf{OpenSSL} and \textbf{Nginx}. OpenSSL  \texttt{s\_server}/\texttt{s\_time} workload collapses in both SYNC and EXITLESS due to Gramine's futex-based file-system synchronisation and the absence of an fd cache, a LibOS-level limitation. In contrast, for Nginx serving a 10~KB HTML file under wrk, fd caching and \texttt{sendfile} allow EXITLESS to achieve a 25\% speedup by avoiding context switches, whereas SYNC loses 50\% over simulation mode.

\textbf{Redis}: the built-in benchmark at 50 concurrent clients and 100K requests per command; EXITLESS loses ${\sim}$10\% to syscall-queue management; SYNC plateaus at ${\sim}$7K ~ops/s (\cref{fig:redis}).  \textbf{Memcached}: \texttt{memtier\_benchmark}~\cite{memtierbenchmark} at 4 threads, 50 connections/thread, 10K requests/client; EXITLESS loses 40\% (higher than Redis due to multi-threaded coordination), SYNC over 80\% (\cref{fig:memcached}). \textbf{MariaDB}: Sysbench~\cite{sysbench} \texttt{oltp\_read\_write} on 10~tables $\times$ 600{,}000~rows at 1--4 threads with the default 128~MB buffer pool; EXITLESS accumulates management overhead with more threads while SYNC loses up to 50\% (\cref{fig:mariadb}).

\subsection{Application Benchmarks with SCONE}\label{real_applications_scone}
We also evaluated \sys under SCONE on Redis, Memcached, and MariaDB to quantify the additional latency introduced by \sys and to compare how different LibOS frameworks shape the overall overhead. We ran the same Redis benchmark and \texttt{memtier\_benchmark} with default parameters, raising \texttt{SCONE\_SSPINS} to 10{,}000 to lengthen the enclave-thread spin window and reduce context switches. As~\cref{fig:scone_apps} shows, \sys adds under 5\% overhead on Redis, essentially matches the baseline on Memcached (HW mode is slightly faster, likely from reduced scheduling overhead), and adds about 12\% on MariaDB.

\subsection{Discussion}\label{evaluation_conclusion}
VMPL transitions are the dominant source of \sys overhead. This is largely invisible for compute-intensive workloads, where the boundary cost is a small fraction of total execution time, but becomes significant for I/O-intensive workloads with frequent syscalls. A well implemented shared-syscall-buffer LibOS (as in SCONE, or EXITLESS in Gramine) can substantially mask this cost.

Frequent access to unmapped shared memory and permission changes on already-synchronized shared pages introduce latency by triggering an extra page-table synchronization operation in \sys. Applications designed for SGX, and for \sys, should avoid these patterns.

eSGX exception mechanism reduces minimal overhead to 3.4 \textmu s compared to AEX enclave exit-and-resume cycle in SGX of at least 36.5 \textmu s. The exception table further enables safer and more efficient exception identification. EMODP accelerates page permission changes entirely in-enclave, while ESETUSSA multiplexes enclave application threads over a single TCS through a simple pointer update.
\section{Related Work}

\textbf{Minimizing the CVM TCB.} Gramine-TDX~\cite{kuvaiskii_gramine-tdx_2024} offloads network and filesystem operations to the host via paravirtualized interfaces but still carries a LibOS. Unikernel- and micro-VM-based confidential runtimes, IncognitOS~\cite{duy_incognitos_2025}, KylinX-style dynamic library OSes~\cite{zhang2018kylinx}, and hardened minimal images~\cite{community_confidential_nodate}, shrink the guest at the cost of compatibility. Intra-CVM privilege separation via VMPLs is complementary: Veil~\cite{ahmad_veil_2023}, Cabin~\cite{cabin}, and Wallet~\cite{wallet} carve CVMs into small trustlets that treat the guest OS as untrusted. \sys sits in this family, guest OS is outside the enclave TCB, VMPL-based enclave isolation and page-table based horizontal privilege separation, host-managed resources.

\textbf{VMPL-based SGX on CVMs.} NestedSGX~\cite{wang_road_2025} emulates SGX leaf functions at the highest VMPL but requires modifying every SGX application's untrusted code and exposes no extended semantics. vSGX~\cite{zhao_vsgx_2022} preserves binary compatibility by splitting the application across a normal VM and a CVM, incurring millisecond cross-VM transitions. Veil~\cite{ahmad_veil_2023} defines its own SGX-style ABI at VMPL0 but cannot run unmodified SGX binaries and requires hypervisor changes. \sys is the first VMPL-based monitor to execute unmodified SGX binaries, SDKs, and shielding runtimes (Gramine, SCONE) inside a CVM, and adds extended semantics (eSGX) absent from prior VMPL designs.

\textbf{Userspace SGX binary compatibility.} Haven~\cite{haven_2015}, Occlum~\cite{shen_occlum_2020}, SCONE~\cite{scone}, Panoply~\cite{shinde_panoply_2017}, Gramine~\cite{tsai_graphene-sgx_2017}, and SGX-LKL~\cite{priebe_sgx-lkl_2020} interpose on in-enclave syscalls via LibOSes or shims, improving compatibility at the cost of in-enclave TCB and often bespoke toolchains. \sys preserves this path unchanged: unmodified binaries under SCONE and Gramine execute on \sys without recompilation.

\textbf{Extended enclave primitives.} MPK-based in-process isolation and privilege hierarchies for SGX are explored in EnclaveDom~\cite{melara_enclavedom_2020}, SGXLock~\cite{chen_sgxlock_nodate}, and LightEnclave~\cite{gu_hardware-software_nodate}; Nested Enclave~\cite{park_nested_2020} proposes multi-layered privilege as a hardware extension; RISC-V proposals such as Timber-V~\cite{weiser_timber-v_2019} explore fine-grained isolation with shared memory. Software-only enclave cloning is explored in MigSGX~\cite{nakashima_migsgx_2021} and in sealing-based work~\cite{gu_secure_nodate}, neither has hardware memory sharing. \sys differs by combining horizontal privilege levels, in-enclave exception and syscall handling, direct memory management, and architectural COW cloning, all in a CVM, without HV changes and keeping SGX-binary compatibility.

\section{Conclusion}

\sys uses AMD SEV-SNP VMPLs to build hardware-protected, process-level enclaves inside a guest CVM, closing the gap between SGX-style isolation and CVM flexibility: the guest OS is outside the TCB, enclaves can be backed by arbitrary guest memory, and strict privilege and exception boundaries are preserved while keeping full userspace binary compatibility with existing SGX stacks. Compute- and memory-bound workloads incur under 5\% overhead (nbench averages 4.8\%); I/O-heavy workloads depend on the libOS, Memcached under Gramine SYNC loses over 80\% throughput, EXITLESS ${\sim}$40\%, and under SCONE both Redis and Memcached stay within 5\%. COW Clone enables fast, memory efficient and practical fork for LibOS multiprocessing.

\bibliographystyle{ACM-Reference-Format}
\bibliography{main.bbl}

\appendix

\section{Complete SGX Instruction Catalogue}
\label{app:instructions}

This appendix lists the full set of SGX structures and instructions adapted or introduced by \sys, and the complete instruction-feature matrix summarised in \cref{sec:sgx-binary-compatibility}.

\subsection{Modified and New Structures and Instructions}

The following SGX structures and instructions are adapted to support the Runtime feature:

\noindent\textbf{SECS.} The existing base/size fields now define the user range; new \emph{runtime base} and \emph{runtime size} fields define the runtime range. A new attribute bit enables the Runtime feature.

\noindent\textbf{TCS.} A new \texttt{OUSSA} field records the USSA offset from the enclave base, analogous to \texttt{OSSA}.

\noindent\textbf{GPRSGX.} A new \texttt{SwitchFlag} field restores the correct execution mode on \texttt{ERESUME} after an AEX.

\noindent\textbf{\texttt{EADD}.} Extended with sanity checks that force metadata pages into the runtime range. A new handler-page type is added via \texttt{EADD}; it holds the predefined entry points for exception and syscall handlers, all of which must be non-empty.

\noindent\textbf{\texttt{EENTER}/\texttt{ERESUME}.} Launch-time checks verify that the handler page is present and that the entire SSA range, including the USSA, is valid, since the runtime exception path can advance CSSA without exiting.

New instructions introduced for the Runtime feature:

\noindent\textbf{\texttt{syscall}.} In user mode, \texttt{syscall} switches to runtime mode and dispatches to the corresponding handler-page entry; no context is saved in the USSA.

\noindent\textbf{\texttt{sysret}.} In runtime mode, \texttt{sysret} returns to user mode, jumping to RCX with sanitized flags in R11; no context is restored from the USSA.

\noindent\textbf{\texttt{ERAISE} and \texttt{EraiseInfo}.} \texttt{ERAISE} raises an SGX-supported exception described in \texttt{EraiseInfo} to the guest OS via the AEX flow. \texttt{EraiseInfo} combines the existing \texttt{ExitInfo} (vector, exit type, validity) and \texttt{EXINFO} (extended info, e.g., page-fault details).

\noindent\textbf{\texttt{ESETTIMER}.} Sets the \texttt{TCS} timer count in 100 $\mu$s units with the highest bit requesting a periodic mode. It is cleared by setting the counter to 0.

\noindent\textbf{\texttt{ESETUSSA}.} Resets the USSA to the address in RBX, updating \texttt{OUSSA} in the current TCS after sanity checks.

\noindent\textbf{\texttt{ESWITCH} and \texttt{SwitchFlag}.} \texttt{ESWITCH} consumes a \texttt{SwitchFlag} in RBX to load context from the specified SSA/USSA, switch mode, and decrement CSSA as needed. \texttt{SwitchFlag} contains a \texttt{USER} bit (mode), an \texttt{SSE\_IGNORE} bit (whether to load SSE state), and a 4-bit target SSA index. \texttt{USER} may only be set with target index~0 (the USSA). \texttt{ESWITCH} additionally checks that the \texttt{USER} and \texttt{SSE\_IGNORE} bits in the target GPRSGX match the input flag.

\noindent\textbf{\texttt{EMODP}.} Unifies \texttt{EMODPR} and \texttt{EMODPE} under a single ABI for complete in-enclave permission control over regular pages.

Additional adaptations and instructions support the Clone feature:

\noindent\textbf{SECS.} A new attribute bit enables the Clone feature.

\noindent\textbf{\texttt{EREMOVE}/\texttt{EMODT}.} Because pages can now be shared across enclaves, the caller must pass the target enclave's SECS to disambiguate ownership.

\noindent\textbf{\texttt{ESMPROV}.} The provisioning interface used by PIE to transfer the KBS-provided provisioning blob into the SM.

\noindent\textbf{\texttt{ENCLS} in clone state.} All \texttt{ENCLS} except \texttt{ESYNC} are disabled while an enclave is mid-clone.

\noindent\textbf{\texttt{ECLONE} and \texttt{ECloneMetadata}.} \texttt{ECLONE} initiates cloning by exiting to a target address and populating \texttt{ECloneMetadata} at a shared address. The metadata carries the clone hash, total page count, and metadata-page count, enough to uniquely identify the clone and coordinate with the driver.

\noindent\textbf{\texttt{ECCREATE} and \texttt{ECCreateInfo}.} \texttt{ECCREATE} creates the child enclave by copying the parent SECS and binds parent and child via \texttt{ECCreateInfo}. The child is created only when every non-SECS field in \texttt{ECCreateInfo} matches the parent's metadata.

\noindent\textbf{\texttt{ECLONEINFO} and request page.} \texttt{ECLONEINFO} queries unmapped virtual addresses for a specific metadata-page type (TCS or SSA). \texttt{ECLONEINFORequest} begins the request page, followed by per-mapping entries; the security monitor fills in the virtual addresses and the valid-entry count.

\noindent\textbf{\texttt{ECADD} and request page.} \texttt{ECADD} installs the queried metadata mappings. \texttt{ECLONEINFORequest} is transformed into the request page by filling in physical addresses and bumping the next-entry index.

\noindent\textbf{\texttt{ECINIT}.} Verifies that all metadata pages have been added, initializes the child enclave, and unfreezes the parent. The input is \texttt{ECCreateInfo}, used to validate binding and memory requirements.

\noindent\textbf{\texttt{ECABORT}.} Aborts cloning on a parent enclave identified by its SECS, reverting security-monitor state and unfreezing the parent.

\noindent\textbf{\texttt{ECADDCACHE}.} Adds a cache block (base, size) to the target enclave; pages are validated and will be allocated during COW. Each enclave has at most one cache block, removed by the monitor once consumed.

\noindent\textbf{\texttt{ECCLEARCACHE}.} Removes all unallocated pages from the current cache block.

\noindent\textbf{\texttt{ECSYNC}.} Registers a COW synchronization page for the target enclave; the page is reclaimed by \texttt{EREMOVE} during destruction.

\subsection{Instruction-Feature Matrix}

\begin{table}[htbp]
    \centering
    \caption{SGX instructions supported by \sys across features.}
    \resizebox{\columnwidth}{!}{%
        \begin{tabular}{|c|c|c|c|c|}
            \hline
            \rowcolor{black}
            \color{white} Instruction Group &
            \color{white} Instruction &
            \makecell{\color{white} SGX \\ \color{white} Compatible} &
            \color{white} Runtime &
            \makecell{\color{white} Runtime \\ \color{white} and Clone} \\
            \hline
            \multirow{20}{*}{ENCLS SGX}
            & EADD & \greencheck & \greencheck & \greencheck \\
            & EAUG & \greencheck & \greencheck & \greencheck \\
            & EBLOCK & \redcross & \redcross & \redcross \\
            & ECREATE & \greencheck & \greencheck & \greencheck \\
            & EDBGRD & \greencheck & \greencheck & \greencheck \\
            & EDBGWR & \greencheck & \greencheck & \greencheck \\
            & EEXTEND & \greencheck & \greencheck & \greencheck \\
            & EINIT & \greencheck & \greencheck & \greencheck \\
            & ELDBC & \redcross & \redcross & \redcross \\
            & ELDB & \redcross & \redcross & \redcross \\
            & ELDUC & \redcross & \redcross & \redcross \\
            & ELDU & \redcross & \redcross & \redcross \\
            & EMODPR & \greencheck & \redcross & \redcross \\
            & EMODT & \greencheck & \greencheck & \greencheck \\
            & EPA & \redcross & \redcross & \redcross \\
            & ERDINFO & \redcross & \redcross & \redcross \\
            & EREMOVE & \greencheck & \greencheck & \greencheck \\
            & ETRACKC & \redcross & \redcross & \redcross \\
            & ETRACK & \redcross & \redcross & \redcross \\
            & EWB & \redcross & \redcross & \redcross \\
            \hline
            \multirow{10}{*}{ENCLS \sys}
            & EADDB & \greencheck & \greencheck & \greencheck \\
            & ECABORT & \redcross & \redcross & \greencheck \\
            & ECADD & \redcross & \redcross & \greencheck \\
            & ECADDCACHE & \redcross & \redcross & \greencheck \\
            & ECCLEARCACHE & \redcross & \redcross & \greencheck \\
            & ECLONEINFO & \redcross & \redcross & \greencheck \\
            & ECCREATE & \redcross & \redcross & \greencheck \\
            & ECINIT & \redcross & \redcross & \greencheck \\
            & ECSYNC & \redcross & \redcross & \greencheck \\
            & ESYNC & \greencheck & \greencheck & \greencheck \\
            \hline
            \multirow{9}{*}{ENCLU SGX}
            & EACCEPTCOPY & \redcross & \redcross & \redcross \\
            & EACCEPT & \greencheck & \greencheck & \greencheck \\
            & EDECCSSA & \redcross & \redcross & \redcross \\
            & EENTER & \greencheck & \greencheck & \greencheck \\
            & EEXIT & \greencheck & \greencheck & \greencheck \\
            & EGETKEY & \greencheck & \greencheck & \greencheck \\
            & EMODPE & \greencheck & \redcross & \redcross \\
            & EREPORT & \greencheck & \greencheck & \greencheck \\
            & ERESUME & \greencheck & \greencheck & \greencheck \\
            \hline
            \multirow{6}{*}{ENCLU \sys}
            & ECLONE & \redcross & \redcross & \greencheck \\
            & EMODP & \redcross & \greencheck & \greencheck \\
            & ERAISE & \redcross & \greencheck & \greencheck \\
            & ESETUSSA & \redcross & \greencheck & \greencheck \\
            & ESWITCH & \redcross & \greencheck & \greencheck \\
            & ESMPROV & \greencheck & \redcross & \redcross \\
            \hline
        \end{tabular}%
    }
    \label{tab:supported-instructions}
\end{table}

\section{Page Ownership and ABA Avoidance}
\label{app:aba}

Each EPCM entry records the owner of a physical page (\cref{app:epcm}). Under Clone, an entry can be shared across enclaves, so the security monitor must ensure that \texttt{ENCLS}/\texttt{ENCLU} requests target pages that actually belong to the intended enclave. Requests originate either from the guest OS via \texttt{ENCLS} (outside the enclave) or from enclave code via \texttt{ENCLU} (inside).

\texttt{ENCLU} operations need no ownership check: the running enclave's identity is known and the security monitor maintains its page table, so any mapping inside the enclave's virtual range belongs to that enclave.

\texttt{ENCLS} operations that cross VMPL boundaries do need validation. Since the EPCM entry alone does not uniquely identify the owner under sharing, the caller must pass the target enclave's SECS, and the monitor verifies that the page belongs to that enclave or one of its descendants.

Because page ownership now changes only on page add or COW with refcount~1, SGX's original assumption, that all enclave pages are removed before the SECS, no longer holds. As a result, the SECS page can no longer safely serve as the enclave identity for cross-enclave ownership tracking.

\begin{figure}[h!]
    \centering
    \includegraphics[width=0.38\textwidth]{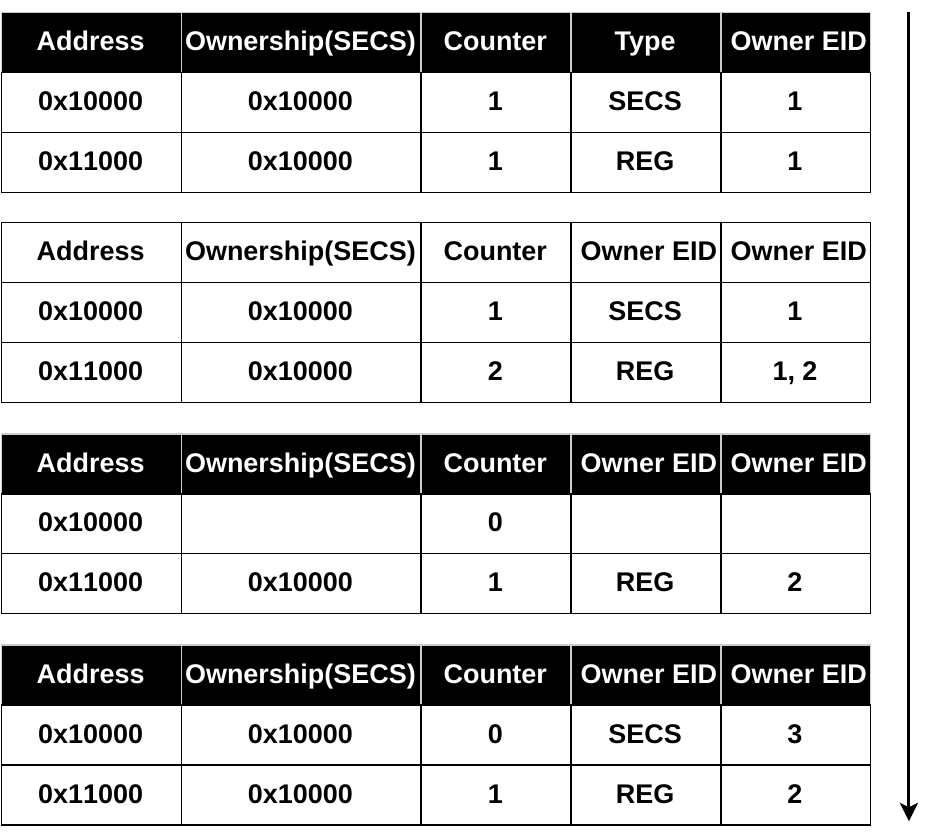}
    \caption{SECS ownership ABA problem.}
    \label{fig:ownership_aba_problem}
\end{figure}

\Cref{fig:ownership_aba_problem} illustrates the failure: an enclave creates a child, the child never writes the shared pages (so COW never runs), and the parent is destroyed. The EPCM still lists the parent as owner. If a new enclave reuses the same physical page for its SECS, the EPCM misleadingly associates the shared pages with a different enclave, an ABA-style ownership confusion that breaks the SGX identity model under sharing.

\begin{figure}[h!]
    \centering
    \includegraphics[width=0.35\textwidth]{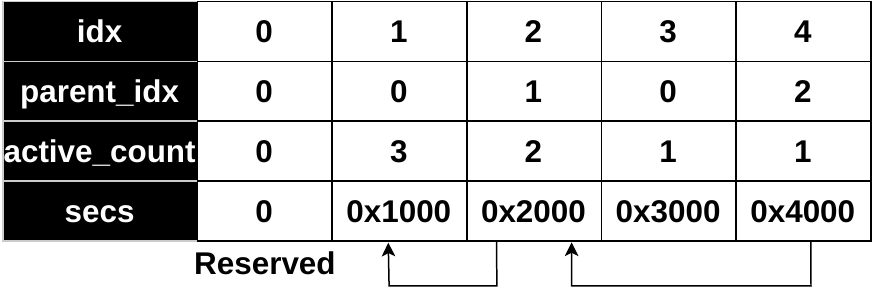}
    \caption{Enclave node ownership list.}
    \label{fig:node_ownership_list}
\end{figure}

To resolve this, \sys explicitly tracks cross-enclave relationships. As~\cref{fig:node_ownership_list} shows, the security monitor maintains an enclave-node list and allocates a free slot per enclave. Each slot stores the parent's index and a counter of active enclaves in its subtree (itself plus all descendants).

On child creation, the monitor increments the counter for the parent and all ancestors; on destruction, it decrements them. The monitor can therefore always tell whether an enclave has active descendants, which suffices for safe ownership validation on EPCM modification and eliminates the ABA confusion above.

\section{Clone/Fork Driver Internals}
\label{app:clone-driver}

\begin{figure}[h!]
    \centering
    \includegraphics[width=0.45\textwidth]{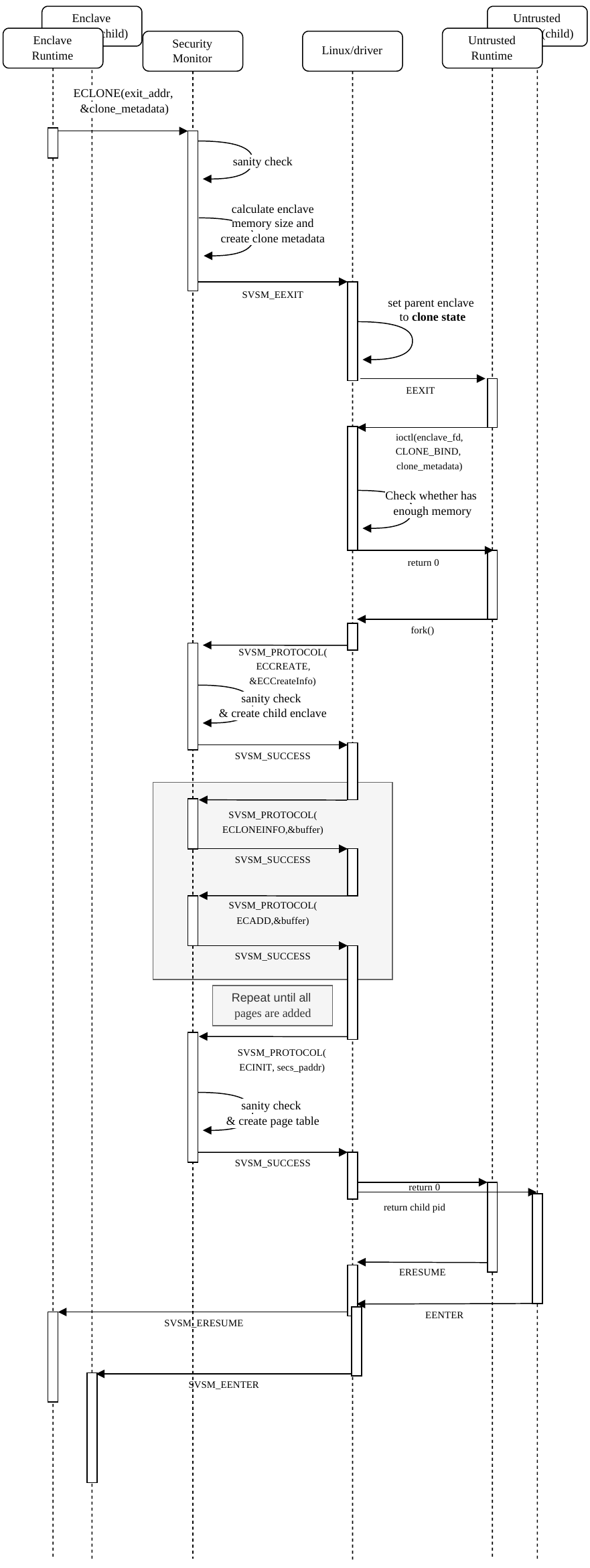}
    \caption{Clone workflow in \sys.}
    \label{fig:clone_sequence}
\end{figure}

During enclave cloning (\cref{fig:clone_sequence}), parent and child processes share the \sys driver through the same file descriptor. The stock SGX driver maps each fd to a single enclave; we extend this to store an xarray in the fd's private data, indexed by address-space pointer. Once a child enclave is created during \texttt{fork}, a new instance is inserted into the xarray under the child's address-space pointer.

To run the entire clone workflow inside \texttt{fork}, \sys hooks the \texttt{vma\_open} callback of the enclave VMA. Before the \texttt{CLONE\_BIND} ioctl returns, the driver allocates a clone-metadata structure and stores a pointer to it in the current enclave. When the kernel duplicates the enclave VMA during \texttt{fork}, the hook runs and starts the clone if the metadata is present and the enclave is in clone state; on failure it invokes \texttt{ECABORT} and frees resources. Because Linux does not check the hook's return value, \texttt{fork} can succeed even when cloning fails, so we add a \texttt{CLONE\_RESULT} ioctl for user space to query the status of the most recent clone.

Each enclave page is described by a two-layer structure in the SGX driver: a page-description structure that records attributes and points to a physical-mapping structure. \sys preserves this layout and extends it for Clone: each enclave keeps its own page-description structure, but multiple enclaves may reference a shared physical-mapping structure whose reference counter gates physical-memory release.

Cache blocks likewise need careful refcounting. Each block is allocated at the maximum size (4~MB) and can only be freed once every page within it has been released. The metadata records a reference counter and a state. Whenever the security monitor allocates a page and publishes it via the synchronization page, the driver creates a physical-mapping structure for the page and increments the block's counter. When a page is released, if it belongs to a cache block the driver only decrements the block's counter; the block is freed only once the counter reaches zero \emph{and} the state is marked freeable. The state becomes freeable either when the counter has reached its maximum (all pages allocated) or when the enclave terminates and the block is removed.

For Clone-enabled enclaves, the driver installs a synchronization page at creation and adds it to an array tracking every synchronization page. A kernel thread periodically walks this array, consuming allocated-page entries and updating the corresponding enclave instances' mappings.

\section{Attestation Protocol Details}
\label{app:attestation}
\subsection{Platform Attestation}
\begin{figure}[h!]
    \centering
    \includegraphics[width=0.46\textwidth]{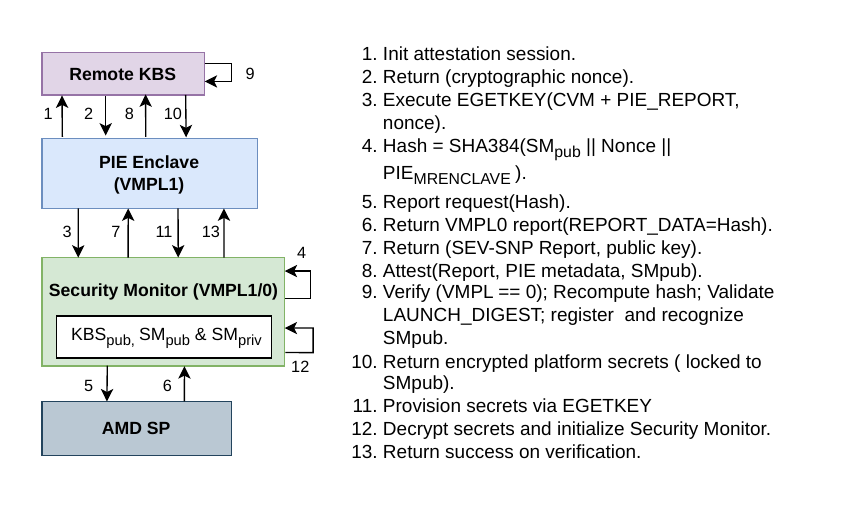}
    \caption{Platform attestation and initialization flow of \sys.}
    \label{fig:teevisor_platform_attestation}
\end{figure}

A platform attestation and initialization service~(PAIS) runs on the guest OS to attest the platform via PIE. The SM recognizes PIE by SIGSTRUCT and grants extended privileged access. PIE obtains a public key and nonce from KBS, requests a VMPL0 report with REPORT\_DATA as hash(SMpub, nonce, MRENCLAVE), and submits both to KBS. On verification, SMpub is registered and KBS returns secrets encrypted with it. The SM validates signatures using KBSpub, confirms correct KBS ownership, and accepts the secrets. All subsequent reports signed by SMpriv are accepted by KBS. \Cref{fig:teevisor_platform_attestation} illustrates this flow.

\subsection{Local Attestation}
Local attestation uses the report key, derived by \texttt{EGETKEY} from ephemeral root keys and additional user input; verification recomputes the MAC over the signed portion of the report and compares it to the MAC field in the report. Because local attestation only verifies locality rather than identity, \sys uses ephemeral root keys generated randomly at bootstrap, every enclave running under the same \sys instance can derive the matching report key.

\subsection{Remote Attestation}
Because a verifiable hardware-embedded key is not directly accessible inside a CVM, \sys bootstraps a monitor-held signing key from the SEV-SNP attestation mechanism in \Cref{app:attestation}. At bootstrap, the monitor generates an attestation key pair (SMpub/SMpriv); after attestation and registration of SMpub a Key Broker Service (KBS). Then, the SMpriv signed report can be verified and accepted by the target KBS.

\section{Instruction-level Latency Breakdowns}
\label{app:latency-breakdown}

\renewcommand{\arraystretch}{1.2}
\begin{table}[htbp]
    \centering
    \caption{Latency of control-handover stages.}
    \label{tab:measurement_stages}

    \begin{minipage}{0.45\columnwidth}
        \centering
        \resizebox{\linewidth}{!}{%
            \begin{tabular}{|c|c|l|}
                \hline
                \rowcolor{black}
                \textbf{\color{white} Instruction} &
                \textbf{\color{white} Stage} &
                \textbf{\color{white} Latency} \\
                \hline
                \multirow{3}{*}{EENTER/ERESUME}
                & 1 & 2.5 $\mu$s \\
                & 2 & 4.5 $\mu$s \\
                & 3 & 2.8 $\mu$s \\
                \hline
                \multirow{3}{*}{EEXIT}
                & 4 & 2.3 $\mu$s \\
                & 5 & 5.4 $\mu$s \\
                & 6 & 2.3 $\mu$s \\
                \hline
                \multirow{2}{*}{AEX}
                & 4 & 1.8 $\mu$s \\
                & 5 & 5.3 $\mu$s \\
                \hline
            \end{tabular}%
        }
    \end{minipage}
    \hfill
    \begin{minipage}{0.48\columnwidth}
        \centering
        \resizebox{\linewidth}{!}{%
            \begin{tabular}{|c|l|}
                \hline
                \rowcolor{black}
                \textbf{\color{white} Stage} & \textbf{\color{white} Description} \\
                \hline
                1 & Untrusted helper code $\rightarrow$ \sys driver end \\
                2 & VMPL2 $\rightarrow$ VMPL1 \\
                3 & Security monitor start $\rightarrow$ enclave \\
                4 & Enclave $\rightarrow$ security monitor end \\
                5 & VMPL1 $\rightarrow$ VMPL2 \\
                6 & \sys driver start $\rightarrow$ untrusted helper code \\
                \hline
            \end{tabular}%
        }
    \end{minipage}
\end{table}

\renewcommand{\arraystretch}{1.2}
\begin{table}[htbp]
    \centering
    \caption{\sys Runtime extension latency.}
    \label{tab:runtime-instructions-latency}
    \resizebox{0.5\columnwidth}{!}{%
        \begin{tabular}{|c|l|l|}
            \hline
            \rowcolor{black}
            \textbf{\color{white} Group} &
            \textbf{\color{white} Instruction} &
            \textbf{\color{white} Latency} \\
            \hline
            \multirow{7}{*}{ENCLU}
            & EMODP & 4.4 $\mu$s \\
            & ERAISE & 7.8 $\mu$s \\
            & ESETUSSA & 3.4 $\mu$s \\
            & ESWITCH & 4.1 $\mu$s \\
            & EXCEPTION & 2.5 $\mu$s \\
            & SYSCALL & 2.3 $\mu$s \\
            & SYSRET & 1.1 $\mu$s \\
            \hline
        \end{tabular}%
    }
\end{table}

\renewcommand{\arraystretch}{1.2}
\begin{table}[htbp]
    \centering
    \caption{\sys Clone extension latency.}
    \label{tab:clone-instructions-latency}
    \resizebox{0.8\columnwidth}{!}{%
        \begin{tabular}{|c|l|l|}
            \hline
            \rowcolor{black}
            \textbf{\color{white} Group} &
            \textbf{\color{white} Instruction} &
            \textbf{\color{white} Latency} \\
            \hline
            \multirow{8}{*}{ENCLS}
            & ECABORT & 16.1 $\mu$s \\
            & ECADD & 14.7 $\mu$s + 1.1 $\mu$s/page \\
            & ECADDCACHE & 343.8 $\mu$s \\
            & ECCLEARCACHE & 386.4 $\mu$s \\
            & ECCREATE & 19.2 $\mu$s \\
            & ECINIT & 114.8 $\mu$s + 8.1 $\mu$s/MB size \\
            & ECLONEINFO & 14.5 $\mu$s \\
            & ECSYNC & 15.0 $\mu$s \\
            \hline
            ENCLU & ECLONE & 12.5 $\mu$s + 0.1 $\mu$s/page \\
            \hline
        \end{tabular}%
    }
\end{table}

\section{Unit Tests}
\label{app:unit-tests}

We ran unit tests to verify instruction correctness, ensuring that the driver and the security monitor take the right execution path along both the native SGX flow and the extended control flow introduced by \sys. Native SGX was validated via the vDSO self-tests, summarized in~\autoref{tab:unit_tests_native}.

\renewcommand{\arraystretch}{1.2}
\begin{table}[htbp]
    \centering
    \caption{\sys unit tests for native SGX.}
    \label{tab:unit_tests_native}
    \resizebox{0.8\columnwidth}{!}{%
        \begin{tabular}{|l|c|}
            \hline
            \rowcolor{black}
            \textbf{\color{white} Description} & \textbf{\color{white} Passed} \\
            \hline
            Debug read/write correctness & \greencheck \\
            PTE permission change and \#PF handling & \greencheck \\
            EPCM permission modification and AEX handling & \greencheck \\
            Dynamic page addition (augment) with EACCEPT & \greencheck \\
            TCS creation/removal and stack/SSA validation & \greencheck \\
            \hline
        \end{tabular}%
    }
\end{table}

\renewcommand{\arraystretch}{1.2}
\begin{table}[htbp]
    \centering
    \caption{\sys unit tests for the Runtime feature.}
    \label{tab:unit_tests_runtime}
    \resizebox{\columnwidth}{!}{%
        \begin{tabular}{|l|l|c|c|}
            \hline
            \rowcolor{black}
            \textbf{\color{white} Instruction} & \textbf{\color{white} Description} & \textbf{\color{white} Expected} & \textbf{\color{white} Passed} \\
            \hline

            \multirow{3}{*}{ECREATE}
            & Runtime size is $<8$ KB or not a power of 2 & \redcross Fail & \greencheck \\
            & Runtime base is lower than user end & \redcross Fail & \greencheck \\
            & Runtime base, size, and features are valid & \greencheck Succeed & \greencheck \\
            \hline

            \multirow{4}{*}{EADD}
            & Handler page has a null entry in handle range & \redcross Fail & \greencheck \\
            & Handler page is not readable/executable & \redcross Fail & \greencheck \\
            & TCS page is not in runtime range & \redcross Fail & \greencheck \\
            & TCS page oentry is not in runtime range & \redcross Fail & \greencheck \\
            \hline

            \multirow{4}{*}{EENTER/ERESUME}
            & SSA or USSA range is not within runtime range & \redcross Fail & \greencheck \\
            & Handler page has not been added & \redcross Fail & \greencheck \\
            & Enclave is actively performing clone & \redcross Fail & \greencheck \\
            & TCS has invalid/missing SSA page & \redcross Fail & \greencheck \\
            \hline

            \multirow{2}{*}{ESWITCH}
            & User switchflag set but target SSA is not 0 & \redcross Fail & \greencheck \\
            & Switchflag in RBX mismatch & \redcross Fail & \greencheck \\
            \hline

            \multirow{2}{*}{ERAISE}
            & Interrupt type is not supported & \redcross Fail & \greencheck \\
            & Eraiseinfo locates outside runtime range & \redcross Fail & \greencheck \\
            \hline

            \multirow{2}{*}{ESETUSSA}
            & USSA range is outside runtime range & \redcross Fail & \greencheck \\
            & USSA range fails SSA sanity check & \redcross Fail & \greencheck \\
            \hline

            EMODPE & Runtime feature is enabled & \redcross Fail & \greencheck \\
            \hline
        \end{tabular}%
    }
\end{table}

The Runtime tests in~\cref{tab:unit_tests_runtime} exercised enclave creation, page additions, SSA/USSA ranges, and control-flow transitions.

To evaluate the Clone feature, the additional test cases in~\autoref{tab:unit_tests_clone} verified the correct behavior of cloning operations.

\renewcommand{\arraystretch}{1.2}
\begin{table}[htbp]
    \centering
    \caption{\sys unit tests for the Clone feature.}
    \label{tab:unit_tests_clone}
    \resizebox{\columnwidth}{!}{%
        \begin{tabular}{|l|l|c|c|}
            \hline
            \rowcolor{black}
            \textbf{\color{white} Instruction} & \textbf{\color{white} Description} & \textbf{\color{white} Expected} & \textbf{\color{white} Passed} \\
            \hline
            ECREATE & Clone enabled but runtime disabled & \redcross Fail & \greencheck \\
            \hline
            EENTER/ERESUME & Clone enabled but sync page not registered & \redcross Fail & \greencheck \\
            \hline
            \multirow{3}{*}{ECLONE}
            & Clone feature is not enabled & \redcross Fail & \greencheck \\
            & Other active TCS exists in enclave & \redcross Fail & \greencheck \\
            & Metadata page not writable/shared & \redcross Fail & \greencheck \\
            \hline
            \multirow{3}{*}{ECCREATE}
            & Info mismatch across pages & \redcross Fail & \greencheck \\
            & Child SECS invalid during init & \redcross Fail & \greencheck \\
            & Info consistency violated & \redcross Fail & \greencheck \\
            \hline
            ECLONEINFO & Enclave is not a child enclave & \redcross Fail & \greencheck \\
            \hline
            ECADD & Metadata type/address incorrect & \redcross Fail & \greencheck \\
            \hline
            \multirow{2}{*}{ECADDCACHE}
            & Enclave does not support clone & \redcross Fail & \greencheck \\
            & Cache blocks already used & \redcross Fail & \greencheck \\
            \hline
            \multirow{2}{*}{ECCLEARCACHE}
            & Cache state is invalid & \redcross Fail & \greencheck \\
            & No cache block exists & \greencheck Succeed & \greencheck \\
            \hline
            \multirow{2}{*}{ECSYNC}
            & Clone feature not enabled & \redcross Fail & \greencheck \\
            & Target page already used & \redcross Fail & \greencheck \\
            \hline
            EAUG, EREMOVE, EMODT & Enclave is performing clone & \redcross Fail & \greencheck \\
            \hline
        \end{tabular}%
    }

    \vspace{0.3cm}

    \resizebox{\columnwidth}{!}{%
        \begin{tabular}{|l|c|}
            \hline
            \rowcolor{black}
            \textbf{\color{white} Description} & \textbf{\color{white} Passed} \\
            \hline
            ECABORT triggers if thread exits between ECLONE and ECINIT & \greencheck \\
            TCS of cloned enclave can enter/resume after clone success & \greencheck \\
            Parent/cloned enclave persists after the other is destroyed & \greencheck \\
            COW triggers on page write or permission modification & \greencheck \\
            \hline
        \end{tabular}%
    }
\end{table}

\section{Clone Performance}
\label{app:clone_performance}
\begin{figure}[htbp]
    \centering
    \includegraphics[width=0.48\textwidth]{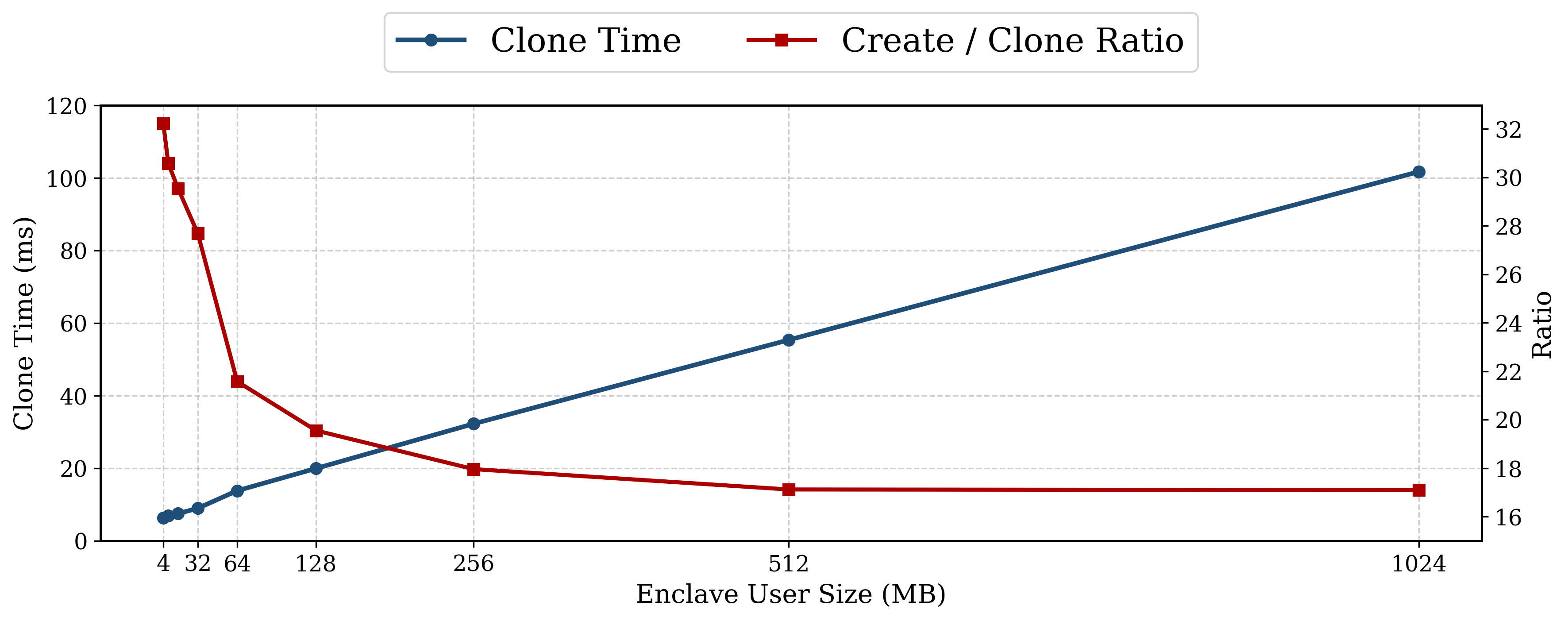}
    \caption{Variation of \sys clone time with a runtime of 64 MB and varying sizes of enclave app, and the associated ratio of enclave creation time to enclave time. The measurements were made at the start and end of the invocation of VMA duplication of the enclave range during fork operation in the kernel.}
    \label{fig:teevisor-clone-performance}
\end{figure}

The cloning time, as measured during duplication of the enclave VMA region by the Linux kernel, during forking, is shown in \cref{fig:teevisor-clone-performance}, showing a gradient of 93$\mu$s per 1~MB increment of enclave size. The enclave creation time at low enclave sizes is dominated by the \sys enclave management metadata and \sys driver's operations in comparison to the same enclave's cloning operations.

\cref{fig:teevisor-clone-performance} shows that fork time scales linearly with the size of the enclave. This dissociates previous SGX forks that scaled with both the size and contents of the enclave.

Minimum enclave time is determined by the size of the enclave metadata (SSA,TCS) and invoked clone instructions.
The total time then scales linearly with enclave size at approximately 93$\mu$s/MB, with most time spent allocating and initializing driver management structures for each enclave page.

\begin{figure}[htbp]
    \centering
    \includegraphics[width=0.48\textwidth]{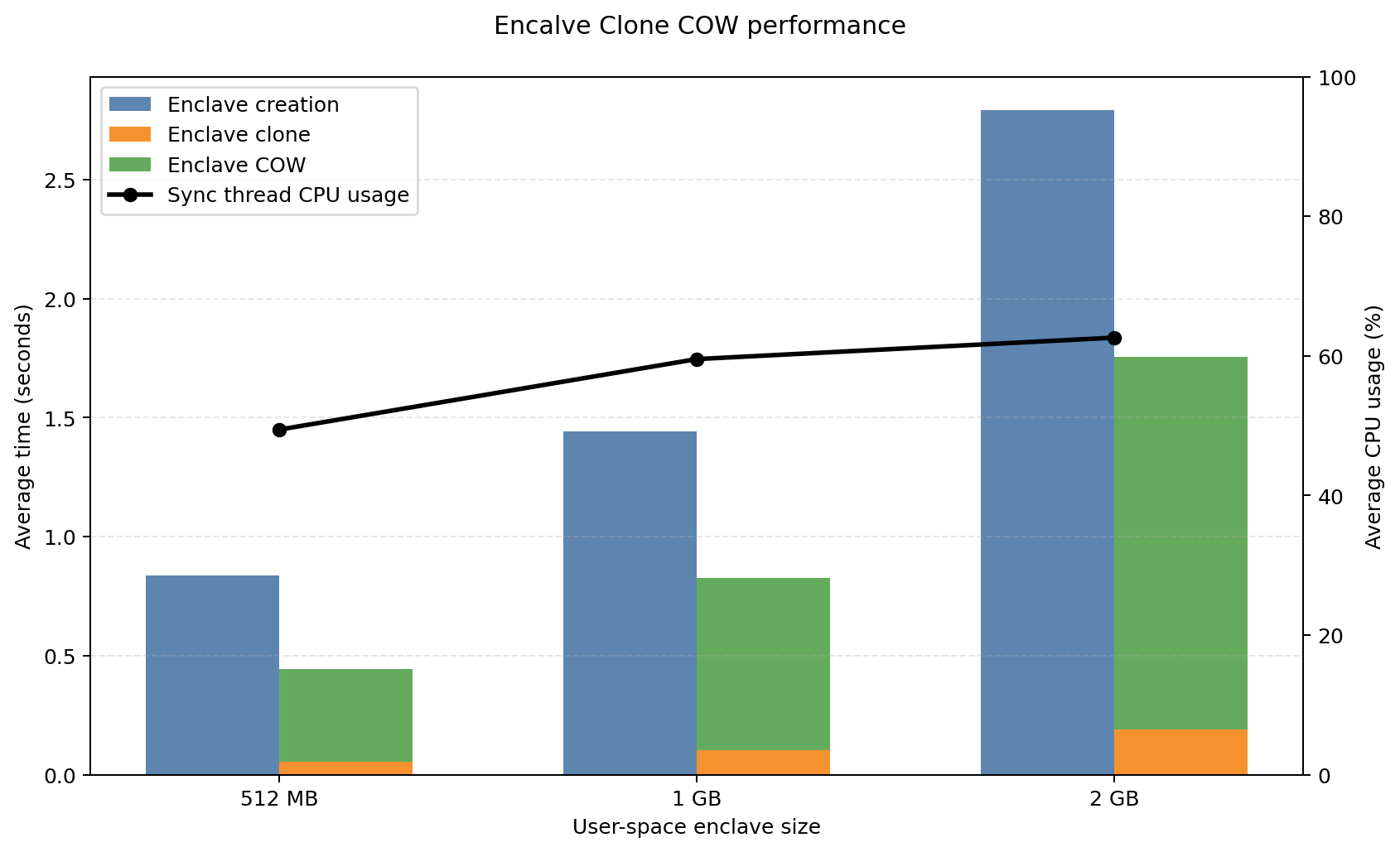}
    \caption{\sys clone COW performance}
    \label{fig:teevisor-cow-performance}
\end{figure}

\cref{fig:teevisor-cow-performance} illustrates the COW performance for user-space enclave sizes ranging from 512~MB to 2~GB and compares the cost of enclave cloning and COW with creating a new enclave in \sys. With the synchronization thread utilizing around 60\% CPU, clone+COW is approximately 1.4--1.7$\times$ faster than creating a new enclave, even without accounting for the additional cost of copying enclave data.

\section{Memory Management}
\label{app:memory_management}

\begin{figure}[h!]
    \centering
    \includegraphics[width=0.45\textwidth]{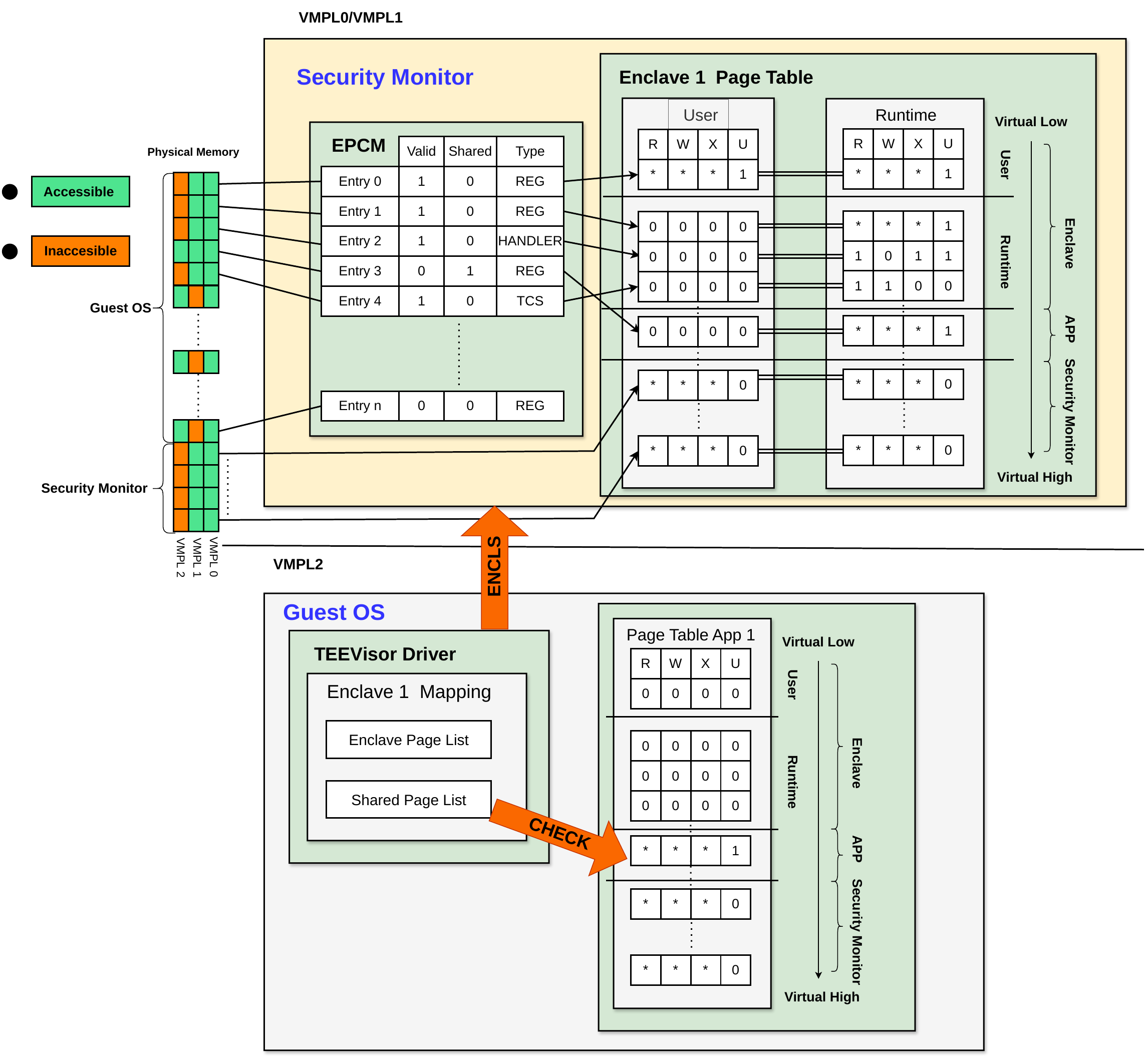}
    \caption{VMPL Layout in \sys.}
    \label{fig:vmpl_layout}
\end{figure}

\Cref{fig:vmpl_layout} shows the layout of physical memory after \sys initialization: \emph{normal memory}, maintained by the guest OS, and \emph{secure memory}, managed by the security monitor. VMPL0 has full access to both; VMPL1 accesses only secure memory; VMPL2 accesses only normal memory. Unlike SGX's fixed EPC, any normal memory page can be converted to enclave memory on demand. The VMPL0 security monitor updates VMPL1 and VMPL2 permissions throughout the enclave life cycle.

\subsection{EPCM and Memory State}
\label{app:epcm}
\label{app:memory_state_update}

\sys extends each EPCM entry with four fields to support memory conversion and sharing: a \emph{shared bit} (since any normal page can back an enclave, a simple valid bit no longer suffices), a \emph{Pvalidate bit} (tracking SEV-SNP acceptance state so the monitor refuses access to unvalidated pages), an 8-bit \emph{reference counter} (for cross-enclave sharing and COW accounting), and a unique \emph{ownership identifier} that replaces the raw SECS reference to avoid ABA confusion after a COW clone.

\sys allocates one EPCM entry per normal-memory page at initialisation, a memory trade-off for flexibility, and grants VMPL1 access to secure memory and VMPL2 to normal memory before yielding to the guest OS. The SVSM page-state-change path is piggybacked to keep EPCM acceptance state in sync with the RMP. During enclave \emph{construction}, the monitor revokes VMPL2 access and grants VMPL1 access, maps each page into the enclave's page table with permissions matching its EPCM entry, builds a second runtime page table when Runtime is enabled, and requires an \texttt{ECSYNC}-registered synchronisation page under Clone. At \emph{execution}, shared-memory mappings are installed lazily via \texttt{ESYNC} during page faults; mode transitions under Runtime switch the active page table; Clone increments shared-page counters on successful child creation and services COW faults from pre-registered cache blocks, with a sole-owner optimisation analogous to Linux COW. At \emph{destruction}, the monitor unmaps and decrements counters; when a page has no remaining users, its contents are cleared, its EPCM reset, VMPL2 access restored, and VMPL1 access revoked.

\end{document}